\documentstyle[prd,aps,epsfig]{revtex}

\begin{document}
%
%
\def\beqa{\begin{eqnarray}}
\def\eeqa{\end{eqnarray}}
\def\be{\begin{equation}}
\def\ee{\end{equation}}
\def \wg#1{\mbox{\boldmath ${#1}$}}
\def \w#1{{\bf #1}}
\draft
\title{Equilibrium sequences of synchronized and irrotational binary systems
	composed of different mass stars in Newtonian gravity}
\author{Keisuke Taniguchi\footnote{Present address: Max-Planck-Institut
f\"ur Gravitationsphysik, Albert-Einstein-Institut,
Am M\"uhlenberg 1, 14476 Golm, Germany; e-mail: {\tt keisuke@aei-potsdam.mpg.de}}
and Eric Gourgoulhon\footnote{e-mail: {\tt Eric.Gourgoulhon@obspm.fr}}}
\address{D\'epartement d'Astrophysique Relativiste et de Cosmologie,
  UMR 8629 du C.N.R.S., Observatoire de Paris, \\
  F-92195 Meudon Cedex, France 
}
\date{23 October 2001}
\maketitle

\begin{abstract}

We study equilibrium sequences of close binary systems on circular orbits and
composed of different mass stars with polytropic equation of state
in Newtonian gravity.
The solving method is a multidomain spectral method
which we have recently developed.
The computations are performed for both cases of synchronized and
irrotational binary systems with adiabatic indices
$\gamma = 3, 2.5, 2.25, 2$ and $1.8$, and for three mass ratios:
$M_1/M_2 =0.5, 0.2$ and $0.1$.
It is found that the equilibrium sequences always terminate at
a mass shedding limit (appearance of a cusp on the surface of the less
massive star). For synchronized binaries,
this contrasts with the equal mass case, where the sequences terminate
instead by a contact configuration.
Regarding the turning point of the total angular momentum (or total energy)
along a sequence, we find that it is difficult to get it
for small mass ratios.
Indeed, we do not find any turning points for $M_1/M_2 \le 0.5$
in the irrotational case.
However, in the synchronized case, it becomes easier again to find one
for mass ratios much smaller than $M_1/M_2 \sim 0.2$.

\end{abstract}

\pacs{PACS number(s): 04.25.Dm, 04.40.Dg, 95.30.Lz, 97.80.-d}

\section{Introduction} \label{s:intro}

Among the binary pulsars observed until now
in the Galaxy\cite{ThorsC99,Lorim01},
there exist three double neutron star systems
which are expected to merge within the age of the Universe
\cite{HulseT75,TayloW89,Wolsz91,StairACLNTTW98,DeichK96}.
Such close binary neutron stars are important not only
as a suitable field for testing
the predictions of general relativity\cite{Will81}
but also as one of the promising sources of gravitational waves
for ground based laser interferometers\cite{Thorn95}
and as one of candidates of gamma-ray burst sources\cite{NarayPP92}.

Binary systems of neutron stars evolve due to the emission of gravitational
waves, which makes the system lose energy and shrinks the orbits.
The evolutionary sequence can be separated into three stages.
The first one is the {\em inspiraling stage} in which the orbital separation
is much larger than the radius of a neutron star,
and the point mass post-Newtonian treatment
constitutes an excellent approximation \cite{Blanc01}.
The second one is the {\em intermediate stage} in which the orbital separation
is only  a few times larger than the radius of a neutron star,
so that hydrodynamics as well as general relativity plays
an important role.
In this stage, since the shrinking time of the orbital radius
due to the emission of gravitational waves is
still larger than the orbital period,
it is possible to approximate the state as quasiequilibrium
(see \cite{BonazGM97,Asada98,Shiba98,Teuko98} for the formulation
and \cite{BonazGM99,MarroMW99,UryuE00,UryuSE00,GourGTMB01}
for numerical computations).
The final stage is the {\em merging stage} in which the two stars coalesce
dynamically
\cite{OoharN97,Shiba99a,Shiba99b,Shiba99c,ShibaU00,OoharN99,Suen99,FontMST00}.

The present article is the third of a series \cite{GourGTMB01,TanigGB01}
devoted to the study of the intermediate stage,
i.e. the quasiequilibrium one,
in order to investigate the physical processes around
the innermost stable circular orbit (ISCO)
as well as to prepare initial conditions for computing 	the merging stage.
In Paper I\cite{GourGTMB01} (general relativistic computations with $\gamma=2$)
and Paper II\cite{TanigGB01} (Newtonian computations with various $\gamma$),
we considered binary systems composed of identical neutron stars
for calculational simplicity.
However, the masses of neutron stars which have been measured
by the pulsar timing technique including relativistic effects
are not exactly the same but span some range\cite{ThorsC99}.

In a recent work \cite{UryuSE00}, Uryu, Shibata, \& Eriguchi
have shown that the coalescence of irrotational binary neutron stars
\footnote{It has been shown that the gravitational-radiation
driven evolution is too rapid for the viscous force to synchronize
the spin of each star with the orbit.
Rather the viscosity is negligible and the fluid velocity circulation
is conserved in these systems.
Provided that the initial spins are not in the millisecond regime,
this means that close binary configurations are well approximated by zero
vorticity states \cite{Kocha92,BildsC92}.}
of equal mass
will produce a disk of negligible mass fraction around black hole.
This makes this process unlikely to be a gamma-ray burst source.
As suggested by them, however, the situation may change
if one considers binary neutron stars of unequal mass.

Until now, all computations of {\em irrotational} binaries of fluid stars
have concerned equal mass systems
\cite{BonazGM99,MarroMW99,UryuE00,UryuSE00,GourGTMB01,UryuE98,TanigGB01}
except for the Newtonian semi-analytic solutions
of the ellipsoidal approximation\cite{LaiRS94b}
and the perturbative approach\cite{TanigN00a,TanigN00b}
\footnote{In Taniguchi \& Nakamura's papers \cite{TanigN00a,TanigN00b},
although the derived equations include
the different masses case, the corresponding results are not shown in tables.
However, it is easy to obtain the results for the different masses configurations
by multiplying some functions of the mass ratio.
One can find such functions from the source terms of
ordinary differential equations which are given
in \cite{TanigN00a,TanigN00b}.}.
Regarding {\em synchronized} binaries, computations of
different mass systems have been performed in Newtonian gravity
\cite{HachiE84,RasioS94,RasioS95}
and in the first post-Newtonian regime \cite{FaberRM01}.
In the present paper, we present numerical results
of synchronized and irrotational binary systems
composed of different mass stars with the same polytropic equation of state
in Newtonian gravity.
The solving method is a multidomain spectral method
which we have developed in both general relativity and Newtonian gravity
\cite{BonazGM98,BonazGM99,GourGTMB01,TanigGB01}.
The results of relativistic calculation will be given
in a forthcoming paper\cite{TanigG01}.

The plan of the article is as follows.
We give a short overview of our method in Sec. II.
The numerical results are presented and discussed in Sec. III.
Section IV is then devoted to the summary.

\section{Method}

\subsection{Equations to be solved}

The reader is referred to Sec.~II.A of Paper~II \cite{TanigGB01}
for the complete set
of equations governing perfect fluid Newtonian binary stars.
Here let us simply mention that,
although we consider binary systems composed of
different mass stars, we assume that the two stars obey to the same
polytropic equation of state. This assumption could easily
be relaxed in our numerical code, in order to study
other astrophysical objects, such as
neutron star/white dwarf binary systems.

\subsection{Numerical procedure}

We are using a multidomain spectral method which has been presented
in great details in Paper~I \cite{GourGTMB01}.
For stiff equation of states (adiabatic index  $\gamma > 2$), we
employ the regularization technique described in Paper~II \cite{TanigGB01} in order
to get rid of the infinite gradient of the density field at the surface
of the star and obtain highly accurate results.

The solution is obtained by means of an iterative procedure,
the various steps of which are given in Sec.~II.B of Paper~II.
The modifications of this procedure to take into account that
the two stars are different concern steps (b) and (c), which become
\begin{itemize}
  \item[(b)]
	The separation between the centers of the two stars, $d$, is
	held fixed.
	Here we define the center as the point of the maximum enthalpy
	(or equivalently maximum density).
	As an initial step, we set the $X$ coordinates
	of the two stellar centers as
	\beqa
	  X_{<1>,\rm ini} &=&-{M_2 \over M_1+M_2} d, \\
	  X_{<2>,\rm ini} &=&{M_1 \over M_1+M_2} d,
	\eeqa
	where $M_1$ and $M_2$ denote the gravitational masses
	(or equivalently baryon masses in Newtonian gravity)
	of the two spherically symmetric initial stars.
	Note here that the location of the rotation axis $X_{\rm rot}$,
	which is initially set to $0$, will change.
	This is because during the relaxation of the binary system
	to the equilibrium figure, the stars will be deformed and
	the positions of the centers of masses of each star will change
	(see Sec. IV.D.1 of Paper I for details).
  \item[(c)]
	By setting the central values of the gradient of enthalpy
	to be zero for each star,
	we calculate the location of the rotation axis and
	the orbital angular velocity $\Omega$
	(see Sec. IV.D.2 of Paper I for details).
	After the determination of the rotation axis,
	the stars are translated parallel to the $X$-axis
	in order for the rotation axis to coincide with the origin
	of the coordinate system.
	In short, we set
	\beqa
	  &&X_{<1>, \rm new} =X_{<1>, \rm old} - X_{\rm rot}, \\
	  &&X_{<2>, \rm new} =X_{<2>, \rm old} - X_{\rm rot}, \\
	  &&X_{\rm rot, new} =0.
	\eeqa
	Since this procedure is only the transfer of
	the coordinate location of the binary system,
	it introduces no change in all the other quantities.
\end{itemize}

Numerous tests passed by our numerical codes have been presented in Papers~I and II.
The only additional test we provide here for different mass systems is
to evaluate the error in the virial theorem (see Eq.~(34) of Paper~II)
and to check that it is small. We systematically list the virial error in
all the tables given in this article. For the $\sim 150$ configurations
listed in Tables~\ref{table1} to \ref{table6}, it is always below
$4\times 10^{-5}$, except for 2 configurations, for which it is around
$10^{-4}$.

\section{Results}

By means of the method presented above,
we have computed equilibrium sequences of both
synchronized and irrotational binary systems on circular orbits
in Newtonian gravity.
In the present paper, we consider the case of binary systems composed of
different mass stars but built upon the same polytropic equation of state.
We use 3 domains (one for the fluid interior) for each star
and the following number of spectral coefficients:
$N_r \times N_{\theta} \times N_{\varphi} = 33 \times 25 \times 24$
in each domain.

\subsection{Equilibrium sequences}

The results for evolutionary sequences (constant-mass sequences)
with adiabatic indices $\gamma=3, 2$ and 1.8
are presented in Tables~\ref{table1} -- \ref{table6}
in the cases of synchronized and irrotational binaries
with mass ratios $M_1/M_2=0.5, 0.2$ and 0.1.

In these tables, $d$ denotes the separation between
the centers of the two stars.
Let us recall that we define the {\em center} of a star  as
the point of maximum enthalpy (or equivalently maximum density).
On the other hand, $d_G$ denotes the separation between the centers
of mass of the two stars.
$R_0$, $a_1$, $a_2$, $a_3$, and $a_{1,{\rm opp}}$ are
values relative to star 1 (less massive star) and denote respectively
the radius of a spherical static star of the same mass,
the radius parallel to $x$-axis toward the companion star,
the radius parallel to $y$-axis,
the radius parallel to $z$-axis, and
the radius parallel to $x$-axis in the direction opposite to the companion star.
The $(x, y, z)$ axes are the same as in Fig. 1 of Paper I. In particular,
the stellar centers are located on the $x$-axis and the $z$-axis is
perpendicular to the orbital plane.
The values $\rho_c$ and $\rho_{c0}$ indicate the central density of star 1
and that of a spherical static star of the same mass.
The prime `` ' '' denotes values relative to the companion
star (massive star).
The normalized quantities $\bar{\Omega}$, $\bar{J}$, and $\bar{E}$ are
defined by 
\beqa
  \bar{\Omega} &:=& {\Omega \over (\pi G \rho_0)^{1/2}}, \\
  \bar{J} &:=& {J \over (G M_1^3 R_0)^{1/2}}, \\
  \bar{E} &:=& {E \over G M_1^2/R_0},
\eeqa
where $\Omega$, $J$, and $E$ denote respectively
the orbital angular velocity,
the total angular momentum, and the total energy of the system,
and $\rho_0$ is the averaged density of a spherical static star having the same mass
as star 1:
\be
  \rho_0 :={3 M_1 \over 4 \pi R_0^3}.
\ee
We give also in Tables~\ref{table1} -- \ref{table6} the virial error,
which provides a measure of the relative accuracy of the numerical computation
(see Eq. (111) of Paper I or Eq. (34) of Paper II for the definition).
It has been shown \cite{BonazGM98} by comparison with exact solutions
(Roche ellipsoids), that the virial error indicator is
very well correlated with the numerical error.

Also listed in Tables~\ref{table1} -- \ref{table6} is the ratio
\be
  \chi := {(\partial H/\partial r)_{\rm eq,comp} \over
  		(\partial H/\partial r)_{\rm pole}},
	\label{e:chi}
\ee
where $(\partial H/\partial r)_{\rm eq,comp}$ 
[resp. $(\partial H/\partial r)_{\rm pole}$]
stands for the radial derivative of the enthalpy
at the point on the stellar surface located in the orbital plane and
looking toward the companion star [resp. at the intersection between
the surface and the axis perpendicular to the orbital plane and going 
through the stellar center ($z$ axis)]. 
This quantity is useful because the mass-shedding limit (``Roche limit'')
corresponds to $\chi=0$ (cf. Sec.~IV.E of Paper~I). When $\chi=0$, 
an angular point (cusp) appears at the equator of the star in the
direction to the companion. 

\begin{figure}
\vspace{0.3cm}
  \centerline{ \epsfig{figure=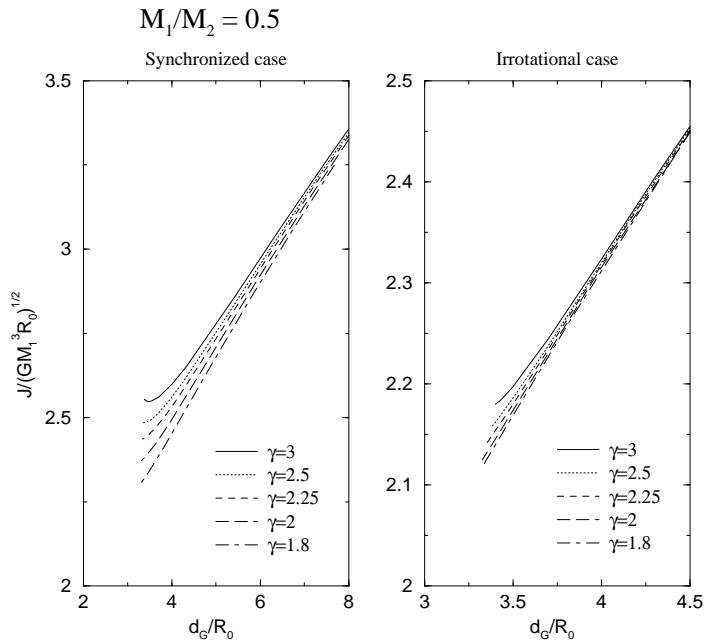,height=8cm} }
\vspace{0.3cm}
\caption[]{\label{fig:angmom_df2}
Total angular momentum along an evolutionary sequence
for the mass ratio $M_1/M_2=0.5$.
The left panel is for synchronized binaries,
and the right one for irrotational binaries. 
Solid, dotted, dashed, long-dashed, and dot-dashed lines denote
the cases of $\gamma=3$, $2.5$, $2.25$, $2$, and $1.8$, respectively.
Note that the scales of the two set of axes are different.}
\end{figure}

\begin{figure}
\vspace{0.3cm}
  \centerline{ \epsfig{figure=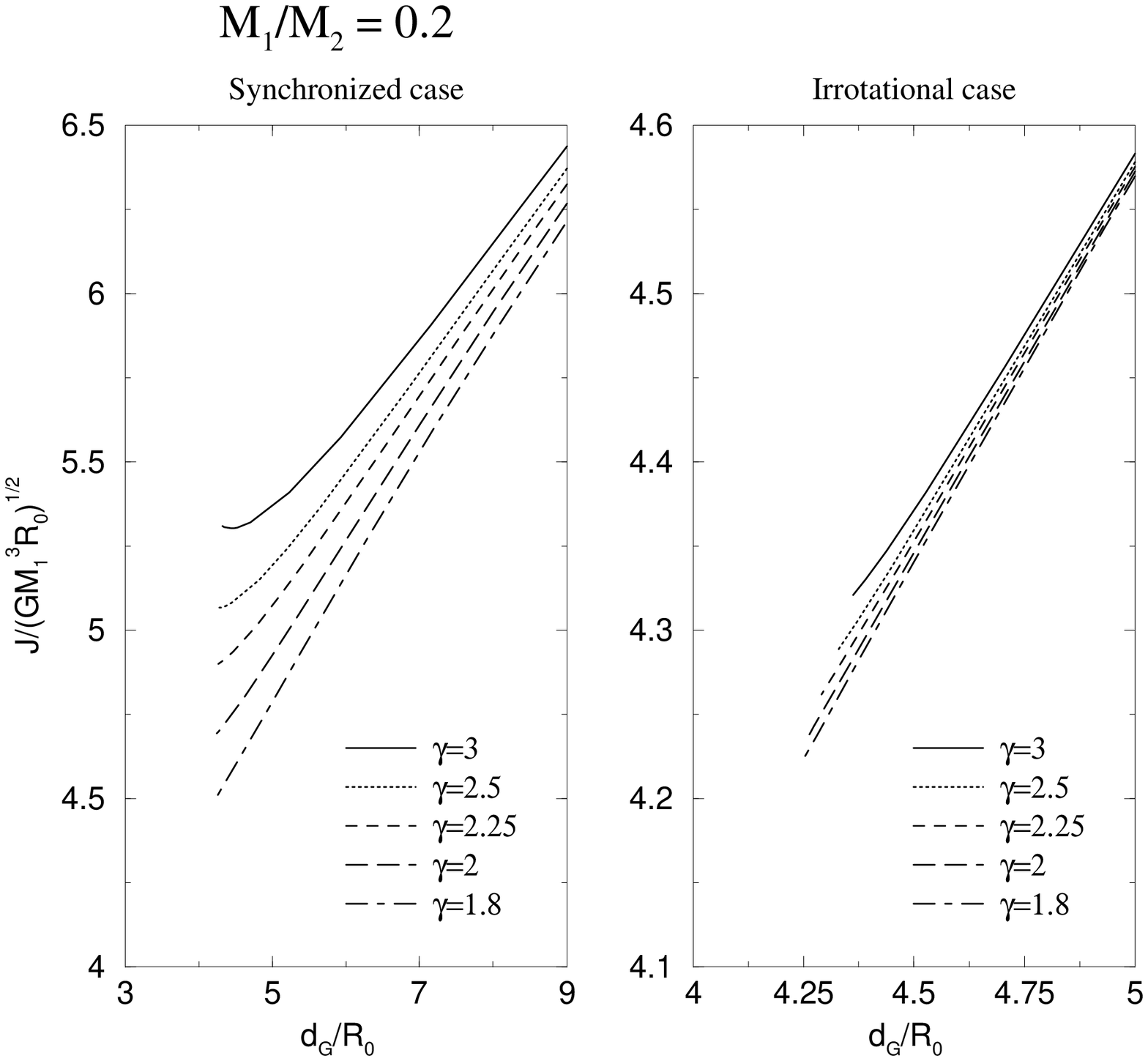,height=8cm} }
\vspace{0.3cm}
\caption[]{\label{fig:angmom_df5}
Same as Fig.~\ref{fig:angmom_df2} but for the mass ratio $M_1/M_2=0.2$.
}
\end{figure}

\begin{figure}
\vspace{0.3cm}
  \centerline{ \epsfig{figure=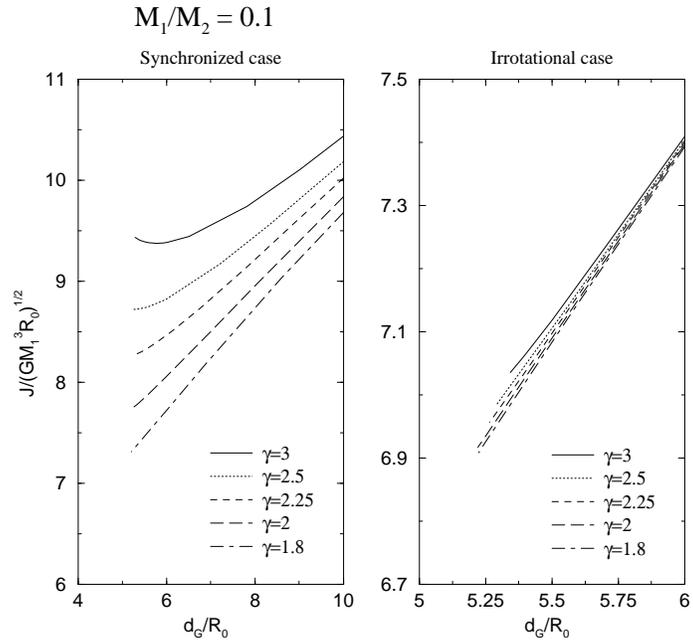,height=8cm} }
\vspace{0.3cm}
\caption[]{\label{fig:angmom_df10}
Same as Fig.~\ref{fig:angmom_df2} but for the mass ratio $M_1/M_2=0.1$.
}
\end{figure}

The total angular momentum along an evolutionary  sequence
is presented in Figs.~\ref{fig:angmom_df2} -- \ref{fig:angmom_df10},
for the mass ratios $M_1/M_2=0.5, 0.2$ and 0.1.
The left panel in each figure is for synchronized binaries,
and the right one for irrotational binaries.
The total angular momenta of irrotational systems
have almost the same values as those of point mass Keplerian motions
because the spin angular momentum of each star is tiny.
On the other hand, the angular momenta of synchronized systems
are larger than those of point mass Keplerian motions
because corotating stars have large spin angular momentum.

\begin{figure}
\vspace{0.3cm}
  \centerline{ \epsfig{figure=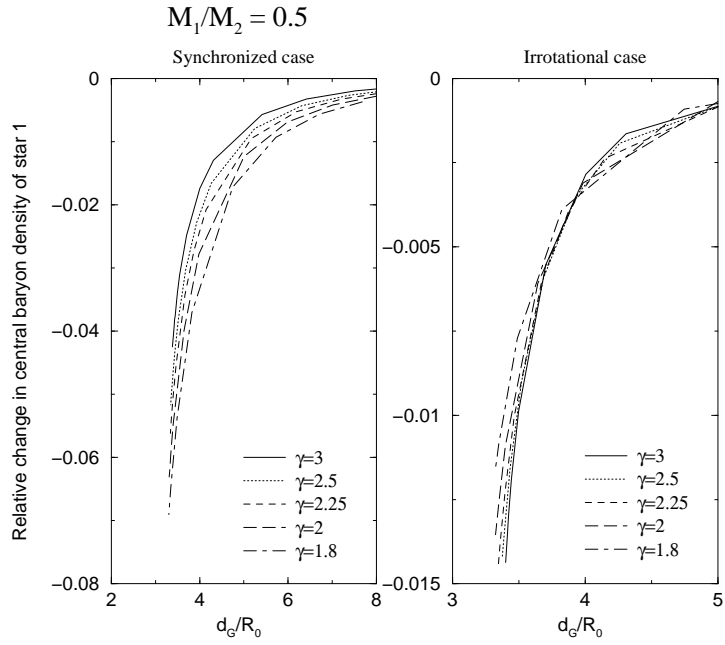,height=8cm} }
\vspace{0.3cm}
\caption[]{\label{fig:change_df2}
Relative change in the central
baryon density of star 1 along an evolutionary sequence for the
mass ratio $M_1/M_2=0.5$. Note that the scales of the two set of axes are different.}
\end{figure}

\begin{figure}
\vspace{0.3cm}
  \centerline{ \epsfig{figure=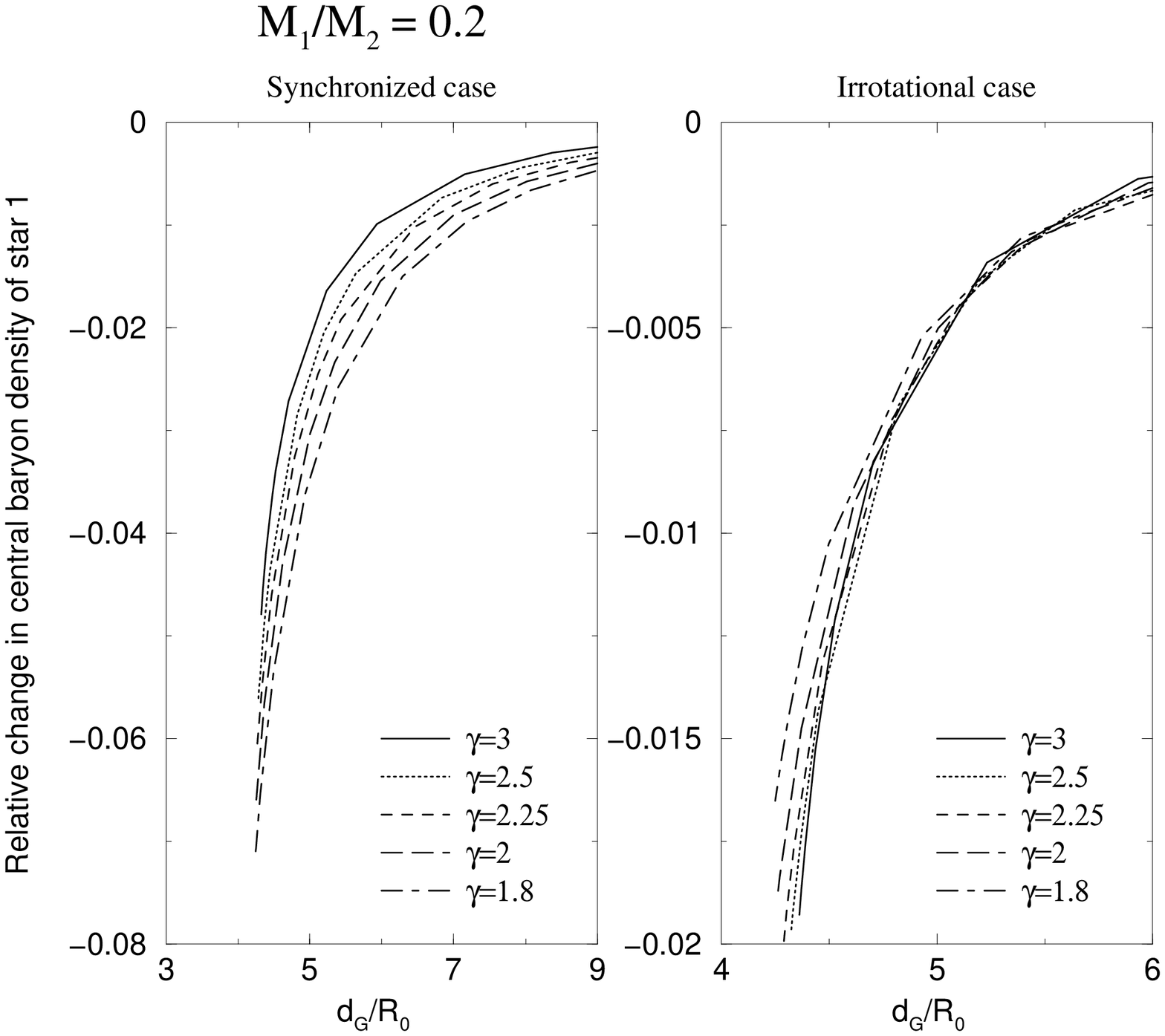,height=8cm} }
\vspace{0.3cm}
\caption[]{\label{fig:change_df5}
Same as Fig.~\ref{fig:change_df2} but for the mass ratio $M_1/M_2=0.2$.
}
\end{figure}

\begin{figure}
\vspace{0.3cm}
  \centerline{ \epsfig{figure=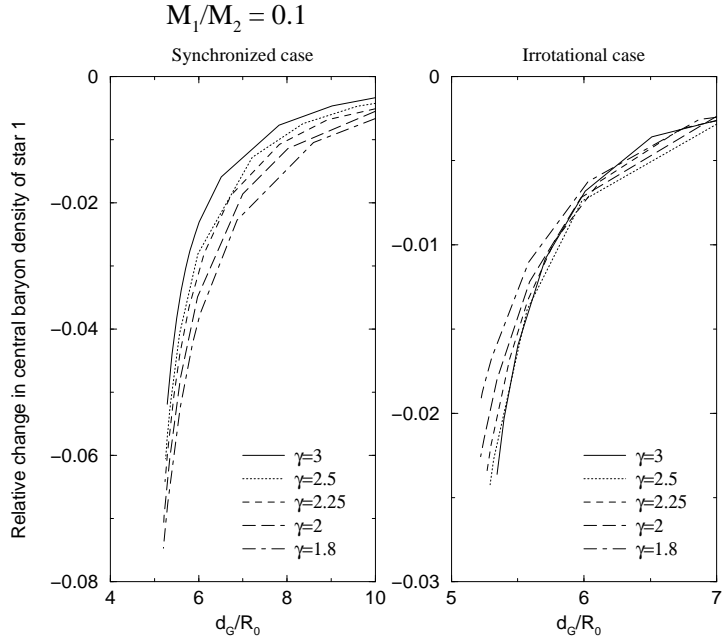,height=8cm} }
\vspace{0.3cm}
\caption[]{\label{fig:change_df10}
Same as Fig.~\ref{fig:change_df2} but for the mass ratio $M_1/M_2=0.1$.
}
\end{figure}

The relative change in central baryon density along an evolutionary
sequence is shown in Figs.~\ref{fig:change_df2} -- \ref{fig:change_df10}
for the mass ratios $M_1/M_2=0.5, 0.2$ and $0.1$.
One can see from these figures that, for all mass ratios,  the
central density decreases monotonically as the two stars come closer.
This behavior agrees with the analytical results
by Taniguchi \& Nakamura\cite{TanigN00a,TanigN00b,TanigN01}.

\begin{figure}
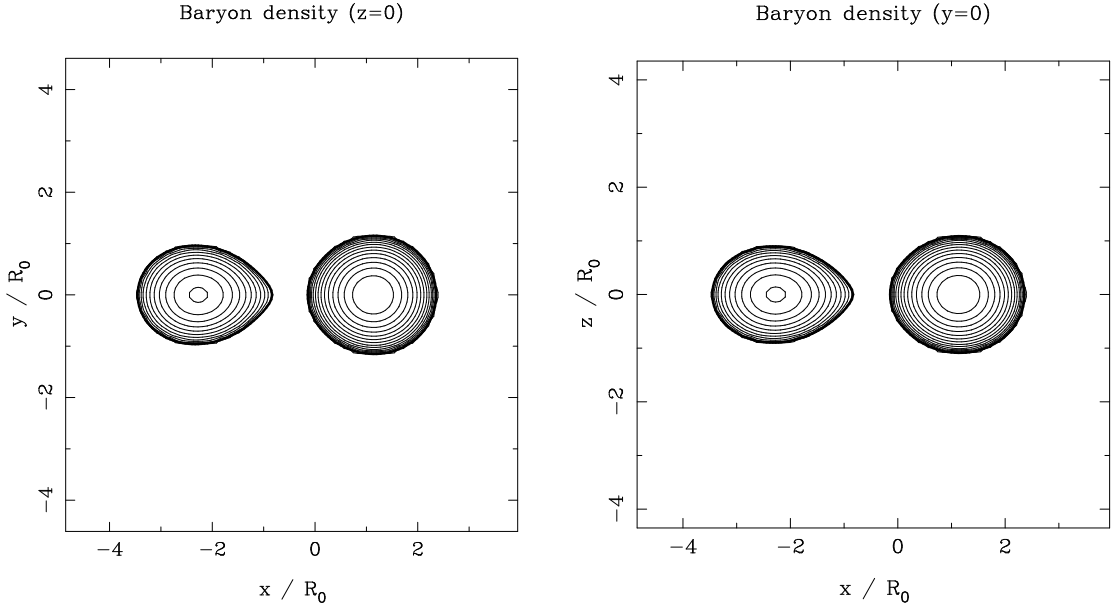

\vspace{0.3cm}
  \centerline{ \epsfig{figure=b_df2xy_Ncg3.eps,height=8cm}
	\hspace{0.5cm} \epsfig{figure=b_df2xz_Ncg3.eps,height=8cm} }
\vspace{0.3cm}
\caption[]{\label{fig:rho_df2_cg3}
Isocontour of the baryon density of synchronized binaries with
$\gamma=3$ and $M_1/M_2=0.5$
when the separation is $d/(R_0+R_0')=1.584$.
The plots are cross sections of $Z=0$ and $Y=0$ planes.
The thick solid lines denote the stellar surface.
The small rough on the stellar surface is an artifact of
the graphical software.
}
\end{figure}

\begin{figure}
\vspace{0.3cm}
  \centerline{ \epsfig{figure=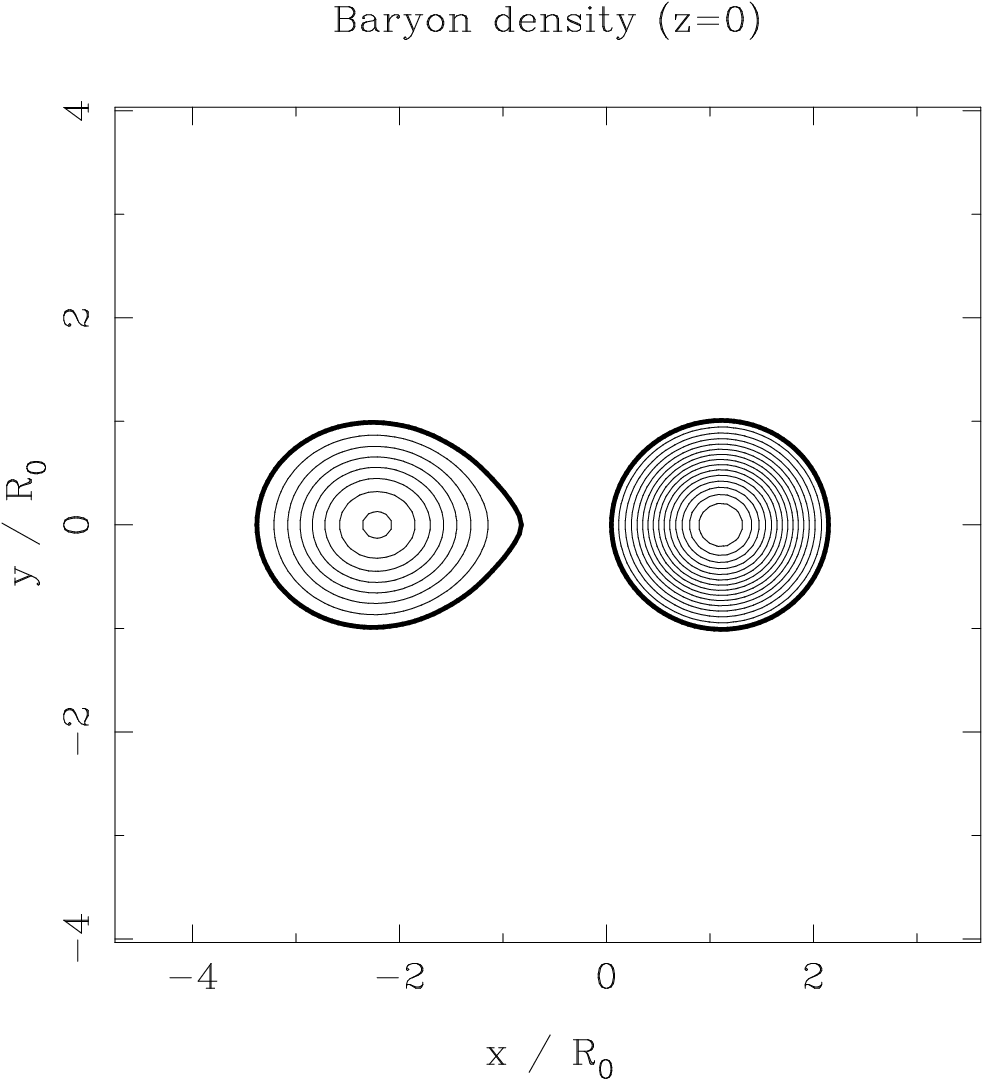,height=8cm}
	\hspace{0.5cm} \epsfig{figure=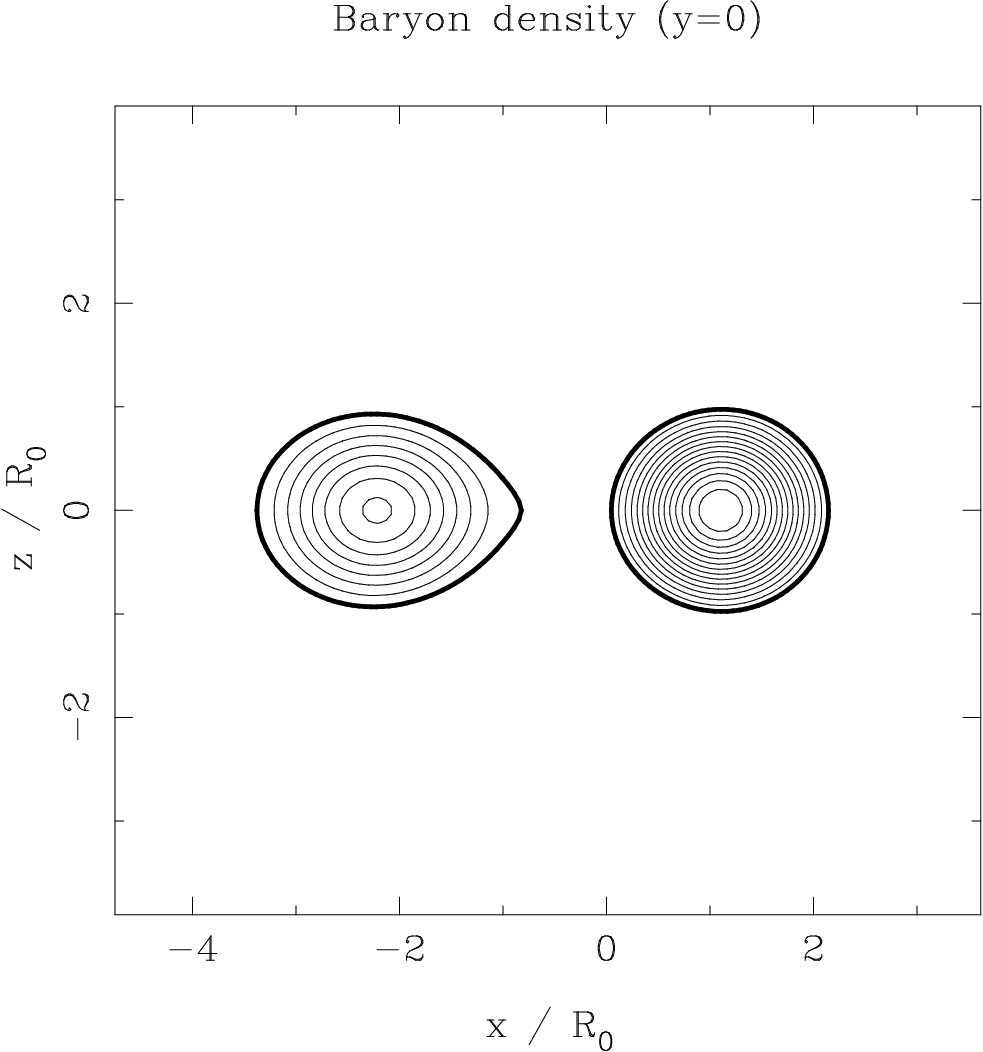,height=8cm} }
\vspace{0.3cm}
\caption[]{\label{fig:rho_df2_cg2}
Same as Fig.~\ref{fig:rho_df2_cg3} but for $\gamma=2$
with the separation $d/(R_0+R_0')=1.660$.
}
\end{figure}

\begin{figure}
\vspace{0.3cm}
  \centerline{ \epsfig{figure=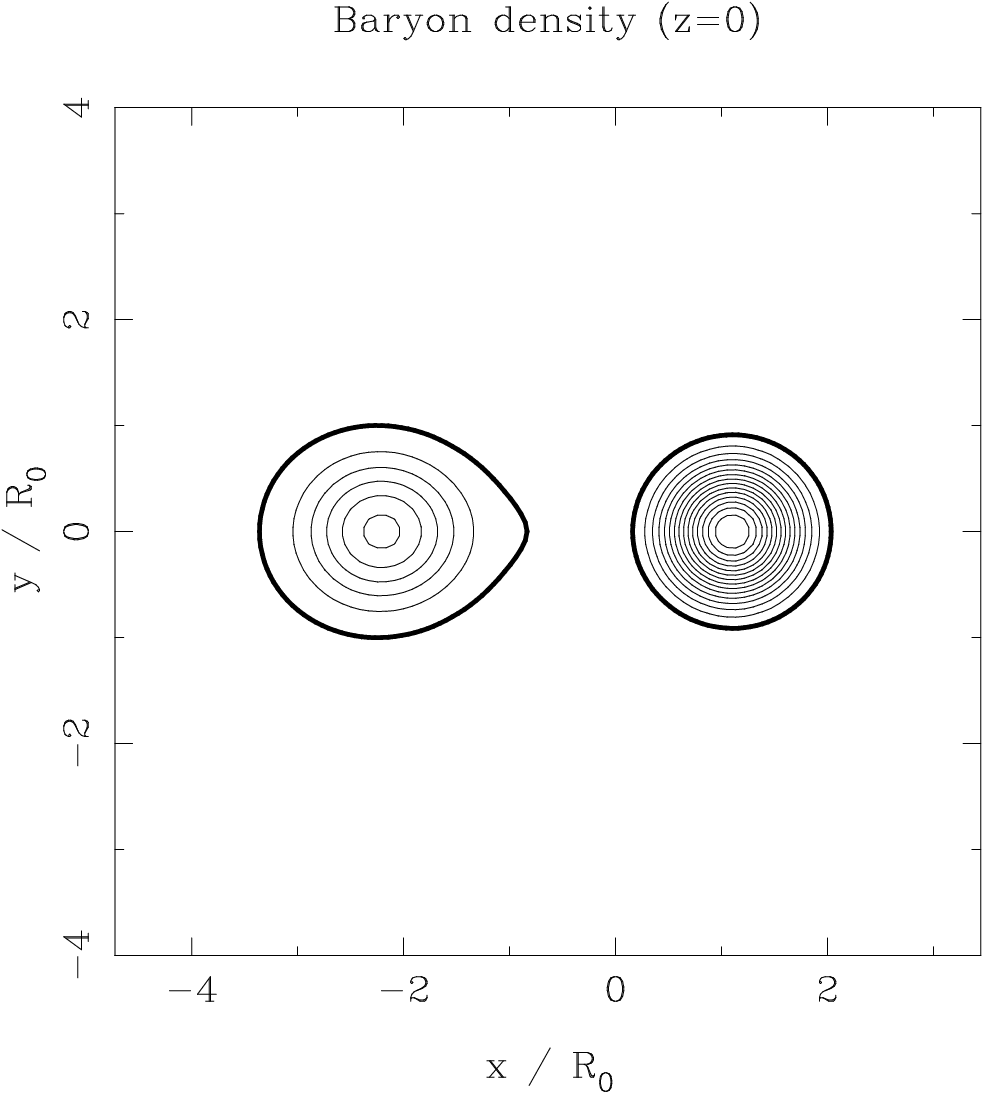,height=8cm}
	\hspace{0.5cm} \epsfig{figure=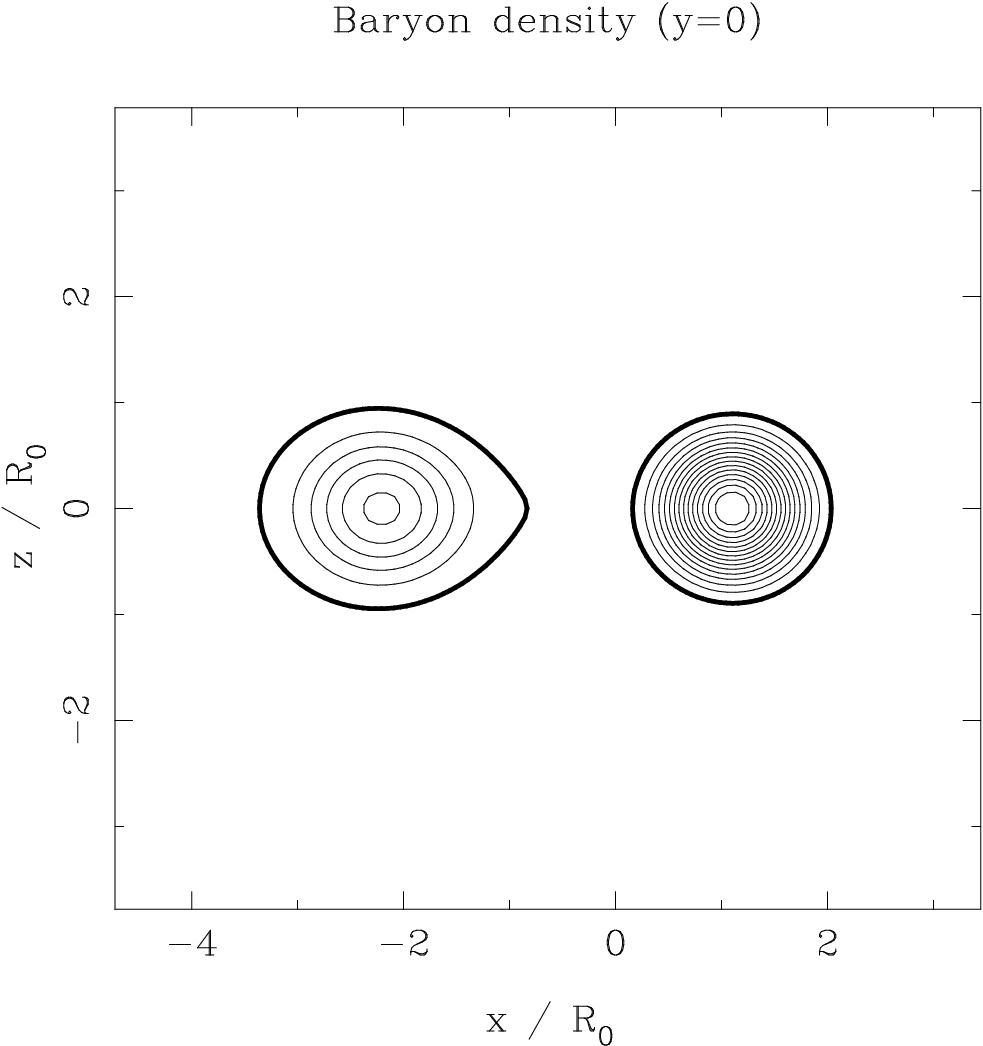,height=8cm} }
\vspace{0.3cm}
\caption[]{\label{fig:rho_df2_cg18}
Same as Fig.~\ref{fig:rho_df2_cg3} but for $\gamma=1.8$
with the separation $d/(R_0+R_0')=1.736$.
}
\end{figure}

\begin{figure}
\vspace{0.3cm}
  \centerline{ \epsfig{figure=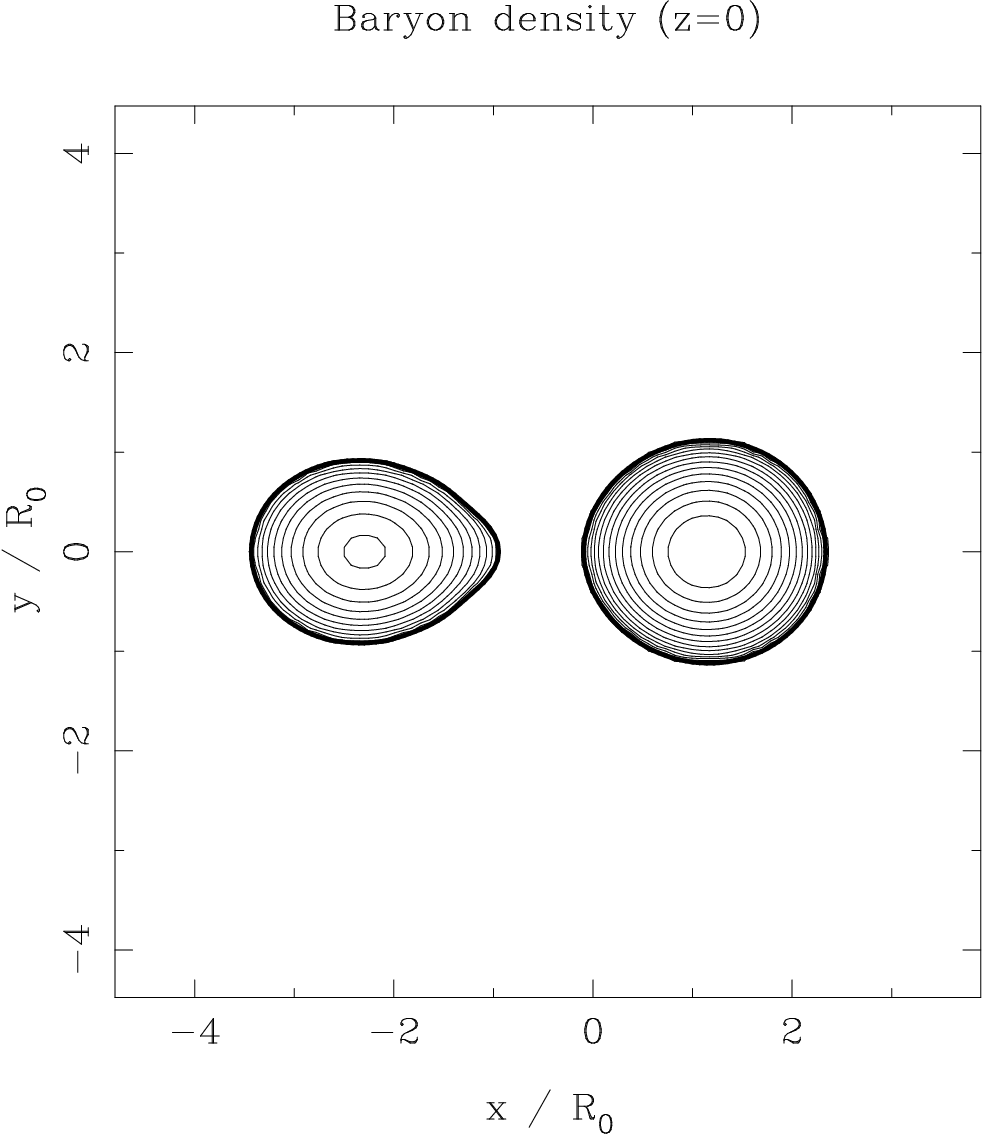,height=8cm}
	\hspace{0.5cm} \epsfig{figure=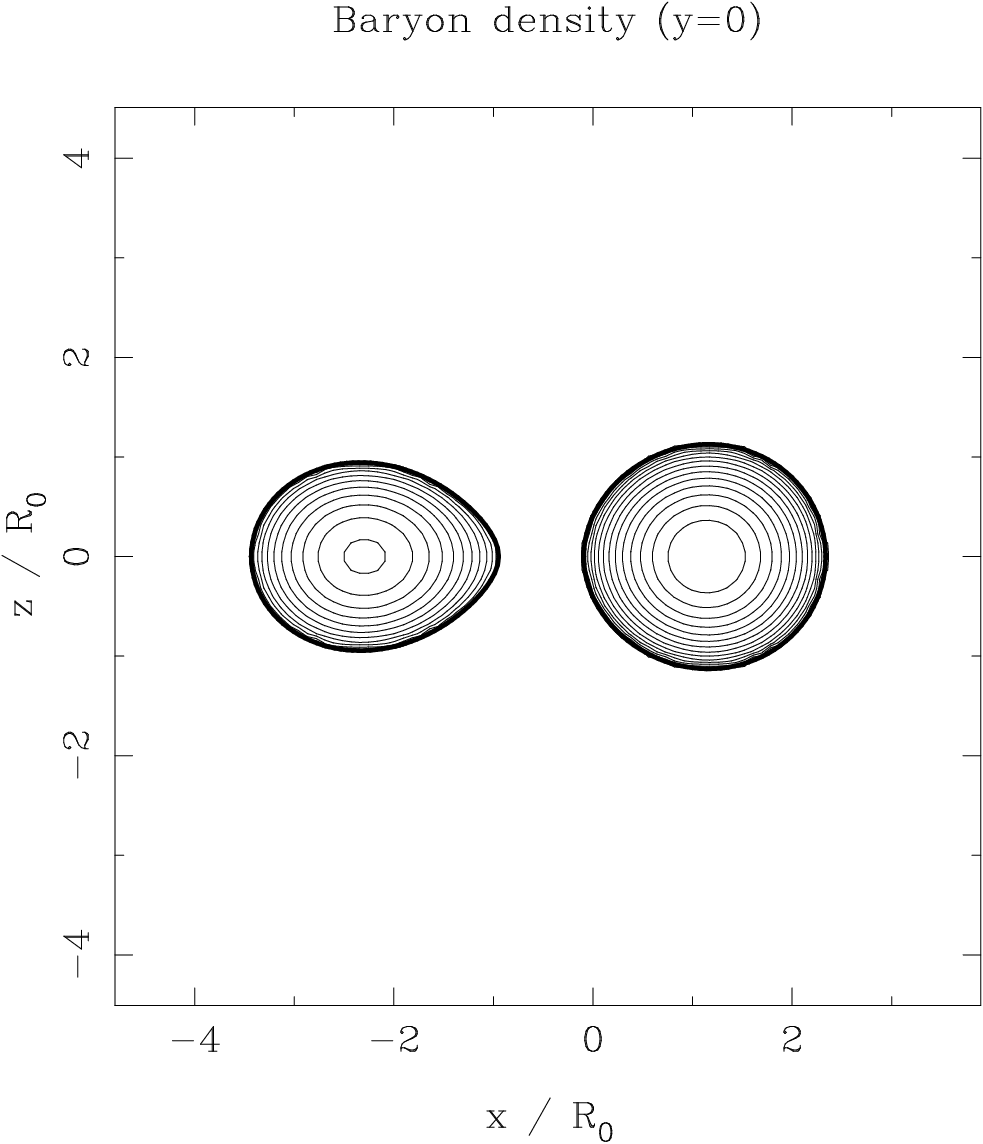,height=8cm} }
\vspace{0.3cm}
\caption[]{\label{fig:rho_df2_ig3}
Isocontour of the baryon density of irrotational binaries with
$\gamma=3$ and $M_1/M_2=0.5$
when the separation is $d/(R_0+R_0')=1.599$.
The plots are cross sections of $Z=0$ and $Y=0$ planes.
The thick solid lines denote the stellar surface.
The small rough on the stellar surface is an artifact of
the graphical software.
}
\end{figure}

Isocontours of the baryon density just before the end point
of equilibrium sequences with $M_1/M_2=0.5$ are presented
in Figs.~\ref{fig:rho_df2_cg3} -- \ref{fig:rho_df2_cg18}
for synchronized binaries and Figs.~\ref{fig:rho_df2_ig3} -- \ref{fig:rho_df2_ig18}
for irrotational binaries.
We can clearly see that there appears a cusp on the surface of
star 1 (less massive star), whereas the system is still well detached.
In the synchronized case, this contrasts with
equal mass binaries, for which we have found that
the equilibrium sequences terminate
by the contact between the two stars (Paper~II).
We will discuss this point in the next section.

\begin{figure}
\vspace{0.3cm}
  \centerline{ \epsfig{figure=b_df2xy_Nig2.eps,height=8cm}
	\hspace{0.5cm} \epsfig{figure=b_df2xz_Nig2.eps,height=8cm} }
\vspace{0.3cm}
\caption[]{\label{fig:rho_df2_ig2}
Same as Fig.~\ref{fig:rho_df2_ig3} but for $\gamma=2$
with the separation $d/(R_0+R_0')=1.677$.
}
\end{figure}

\begin{figure}
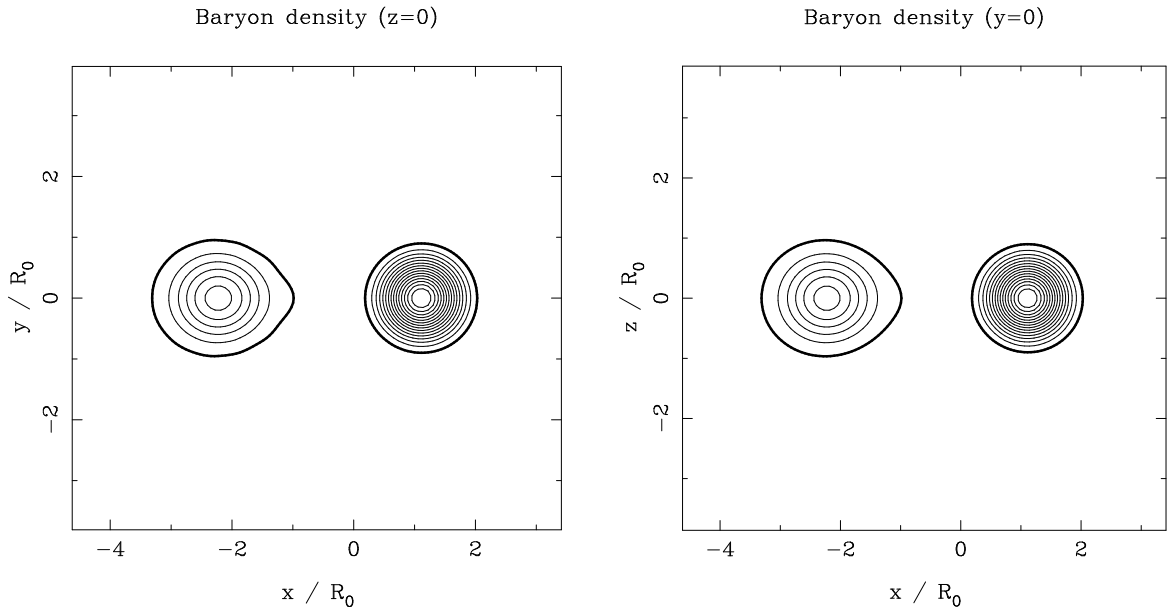

\vspace{0.3cm}
  \centerline{ \epsfig{figure=b_df2xy_Nig18.eps,height=8cm}
	\hspace{0.5cm} \epsfig{figure=b_df2xz_Nig18.eps,height=8cm} }
\vspace{0.3cm}
\caption[]{\label{fig:rho_df2_ig18}
Same as Fig.~\ref{fig:rho_df2_ig3} but for $\gamma=1.8$
with the separation $d/(R_0+R_0')=1.751$.
}
\end{figure}

\subsection{End points of sequences}

A good indicator of the appearance of a cusp at the stellar surface
is the  quantity $\chi$ already introduced in Papers~I and II
and defined by Eq. (\ref{e:chi}).
Comparing $\chi$ with $d/(a_1+a_1')$, which becomes unity
for a contact configuration,
we can determine whether an equilibrium sequence will end by
a mass-shedding configuration or by a contact one.

\begin{figure}
\vspace{0.3cm}
  \centerline{ \epsfig{figure=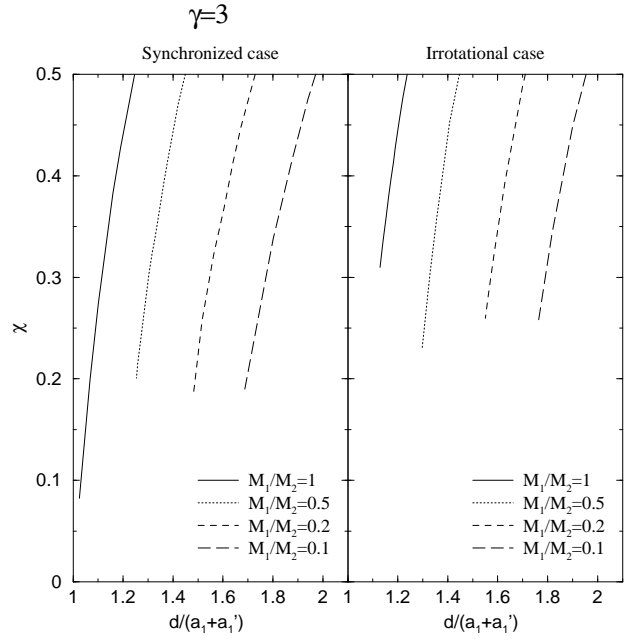,height=8cm} }
\vspace{0.3cm}
\caption[]{\label{fig:chi_g3}
Equatorial to polar ratio of the radial derivative of enthalpy 
$\chi$ as a function of the separation $d$ 
normalized by the total radius $a_1+a_1'$ in the case of $\gamma=3$.
The left (resp. right) panel is for synchronized (resp. irrotational)
binaries. Solid, dotted, dashed, and long-dashed lines denote the mass ratios
$M_1/M_2=1, 0.5, 0.2$, and 0.1, respectively.
}
\end{figure}

\begin{figure}
\vspace{0.3cm}
  \centerline{ \epsfig{figure=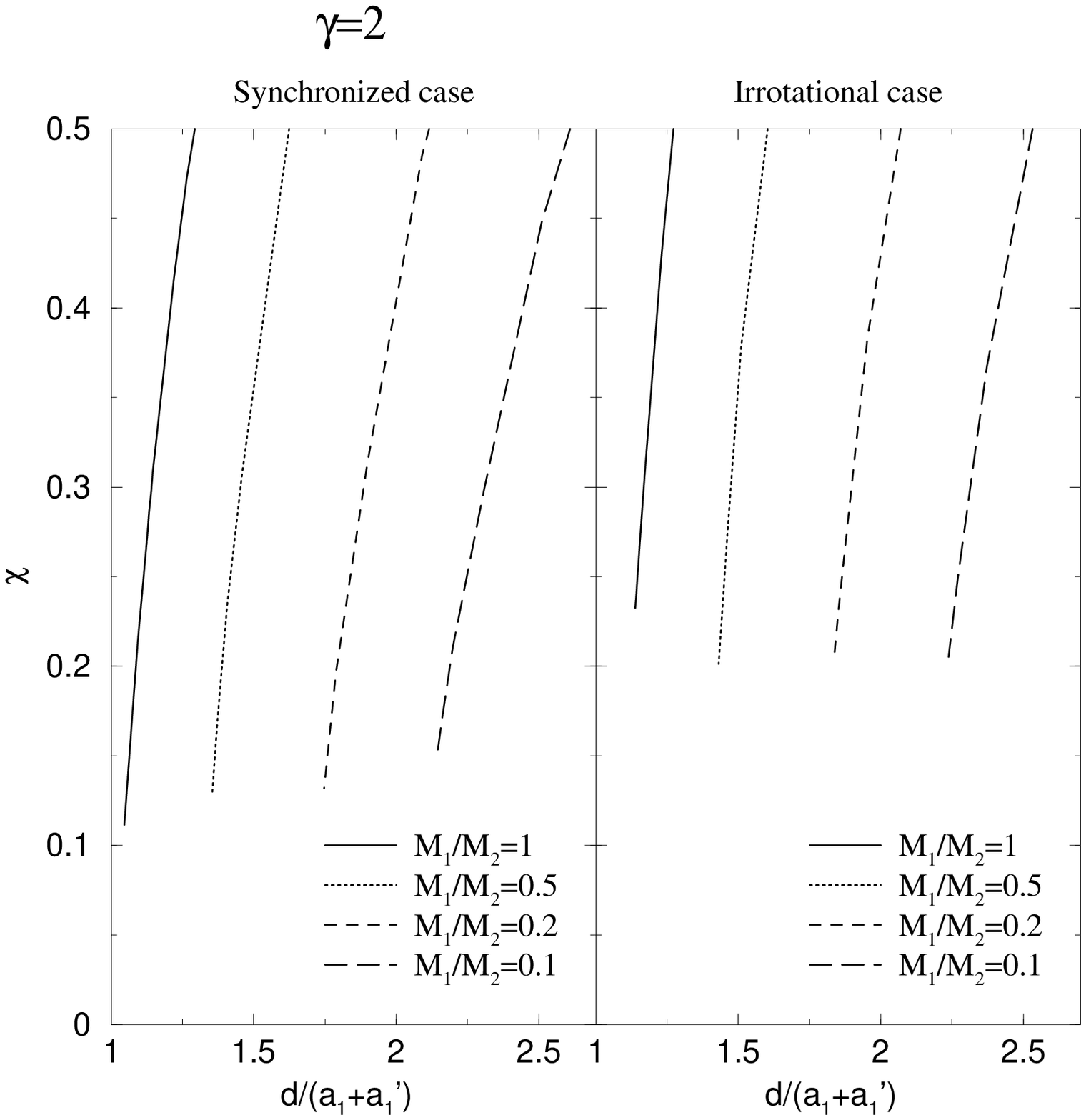,height=8cm} }
\vspace{0.3cm}
\caption[]{\label{fig:chi_g2}
Same as Fig.~\ref{fig:chi_g3} but for $\gamma=2$.
}
\end{figure}

\begin{figure}
\vspace{0.3cm}
  \centerline{ \epsfig{figure=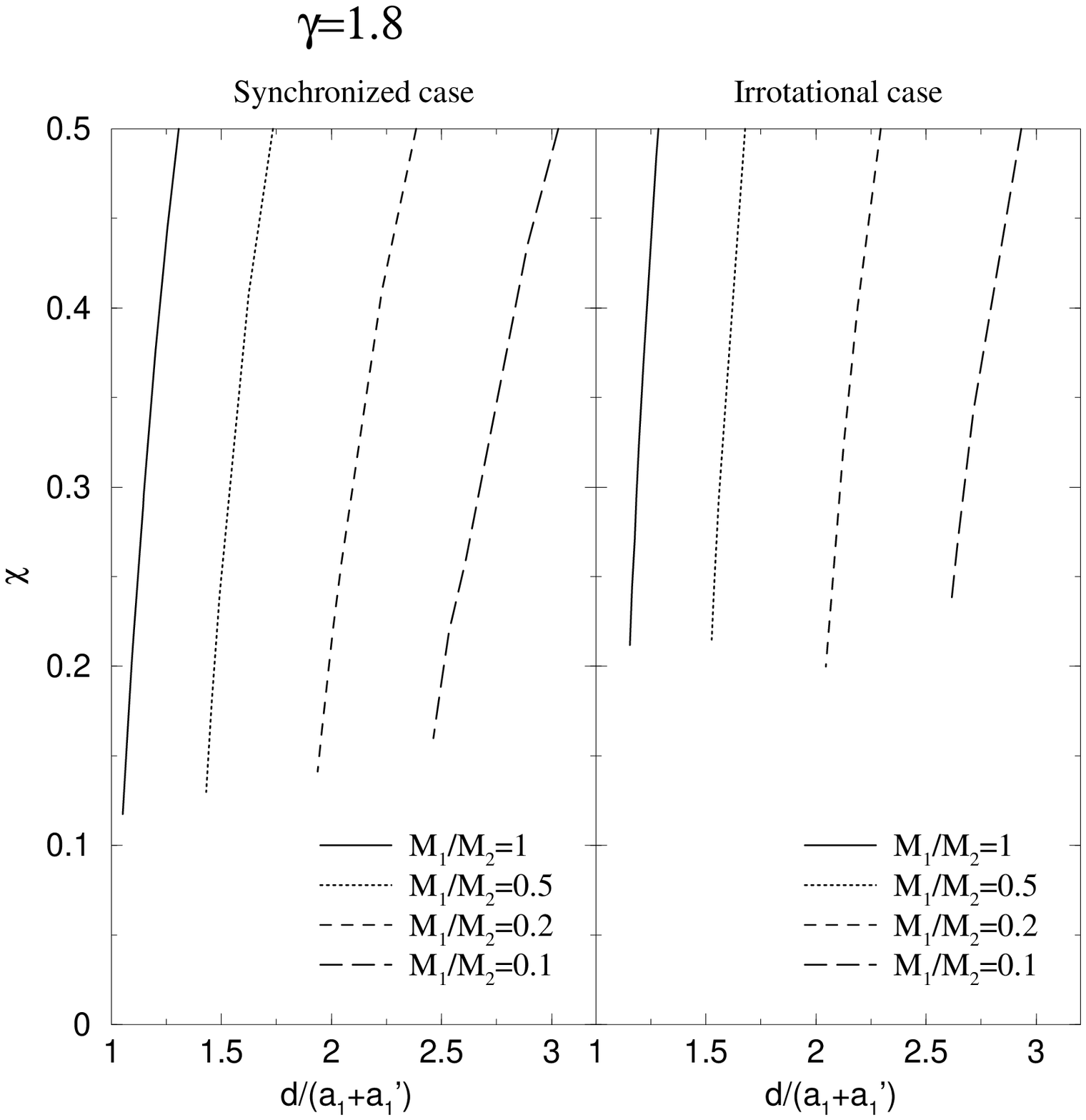,height=8cm} }
\vspace{0.3cm}
\caption[]{\label{fig:chi_g18}
Same as Fig.~\ref{fig:chi_g3} but for $\gamma=1.8$.
}
\end{figure}

\begin{figure}
\vspace{0.3cm}
  \centerline{ \epsfig{figure=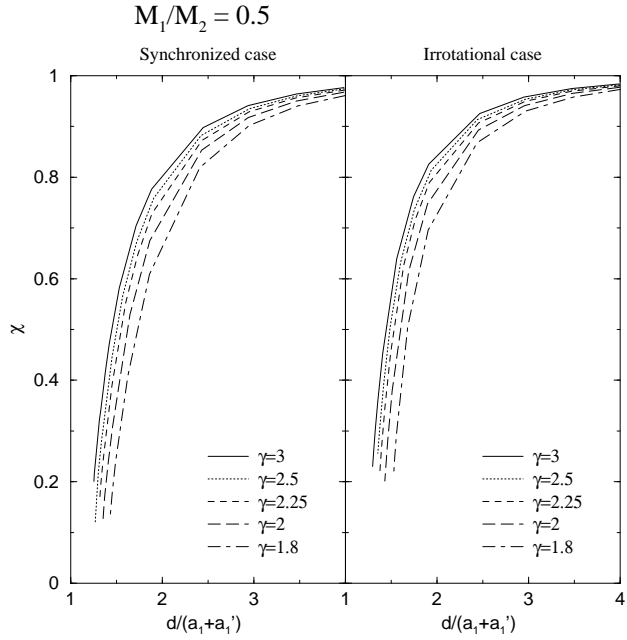,height=8cm} }
\vspace{0.3cm}
\caption[]{\label{fig:chi_df2}
Equatorial to polar ratio of the radial derivative of enthalpy 
$\chi$ as a function of the separation $d$ 
normalized by the total radius $a_1+a_1'$. The mass ratio is $M_1/M_2=0.5$.
The left (resp. right) panel is for synchronized (resp. irrotational)
binaries. Solid, dotted, dashed, long-dashed, and dot-dashed lines denote
the cases of $\gamma=3, 2.5, 2.25, 2$, and 1.8, respectively.
}
\end{figure}

In Figs.~\ref{fig:chi_g3} -- \ref{fig:chi_g18},
the quantity $\chi$ is presented as a function of
the normalized separation $d/(a_1+a_1')$ for $\gamma=3, 2$ and 1.8.
In each figure, we show synchronized and irrotational configurations
with the mass ratios $M_1/M_2=1, 0.5, 0.2$ and 0.1.
Another view of the quantity $\chi$ is presented in Fig.~\ref{fig:chi_df2}.
In this figure, we fix the mass ratio to 0.5 and change $\gamma$.
By extrapolating the curves toward $\chi=0$ in the figures,
we find that all the constant mass sequences terminate by a mass-shedding point
which corresponds to a detached configuration ($d/(a_1+a_1')>1$),
except for the synchronized case with $M_1/M_2=1$. For irrotational
binaries with $M_1/M_2=1$, the detached final configuration seems
marginal from Figs.~\ref{fig:chi_g3} -- \ref{fig:chi_g18}.
However, as discussed in more details Paper~II, it seems that it is
really the case, at least for small values of $\gamma$.

It would not be necessary to extrapolate the curves
of Figs.~\ref{fig:chi_g3} -- \ref{fig:chi_df2}
if we could compute up to $\chi=0$ (cusp).
However, when a cusp appears the stellar surface is no longer
differentiable. Now our multidomain method makes the boundary of the inner domain
fit with the stellar surface \cite{BonazGM98}.
This procedure is essential to get accurate solutions in the irrotational case
(see Appendix~B of Paper~I),
and also necessary for the regularization of the density profile when $\gamma > 2$
(see Sec.~III of Paper~II).
When fitting the inner domain boundary to the stellar surface, the
latter is assumed to be differentiable, being described by a finite
series of differentiable functions of $(\theta,\varphi)$.
This explains why we cannot treat cuspy surfaces.
Furthermore, for very close configurations,
just prior to the apparition of the cusp,
the surface is highly distorted so that there appear unphysical
oscillations when using a finite series of differentiable functions
(Gibbs phenomenon).

\subsection{Turning points of the total angular momentum} \label{s:turning}

In this section, we discuss the turning point of the total
angular momentum (and/or total energy\footnote{The turning points of the total energy
and the total angular momentum always coincide
($dE = \Omega dJ$)\cite{LaiRS93,OstriG69}}) along an evolutionary sequence.
This turning point is interesting
because it corresponds to the onset of the dynamical instability
in the irrotational case which is regarded as a realistic rotational state
of coalescing binary neutron stars \cite{LaiRS94a},
while it signals the secular instability in the synchronized case \cite{BaumgCSST98}.

As one can see from Tables~\ref{table1} -- \ref{table3} and
Table I of Paper II,
there always appears a turning point in the $\gamma=3$ case
for synchronized binaries.
On the other hand, we do not find any turning point
in the irrotational cases for different mass binaries.
Of course, there exists the possibility that the turning point is located
just before the apparition of a cusp for slightly different mass binaries.
As mentioned above, it is difficult for our method to accurately calculate
just before the apparition of a cusp because of the Gibbs phenomenon.
Therefore, we will discuss the appearance of
the turning point only in the synchronized case.

\begin{figure}
\vspace{0.3cm}
  \centerline{ \epsfig{figure=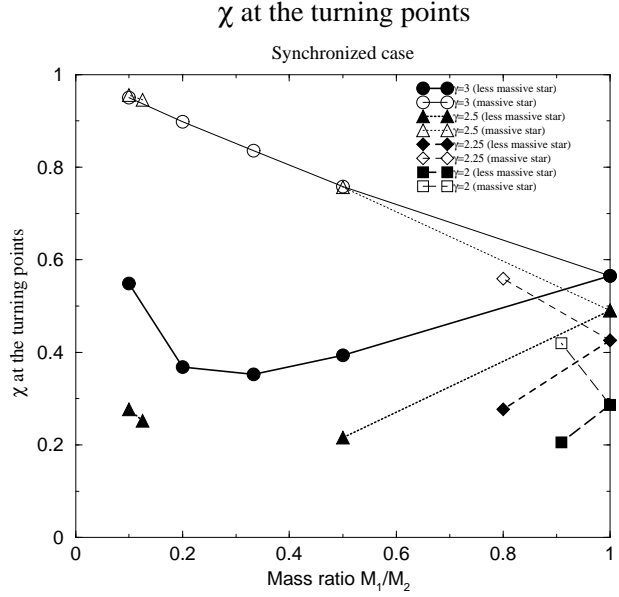,height=8cm} }
\vspace{0.3cm}
\caption[]{\label{fig:turning}
Quantity $\chi$ at the turning points of the total angular momentum
(and/or the total energy)
as a function of the mass ratio $M_1/M_2$ in the synchronized case.
Solid lines with circles, dotted with triangles,
dashed with diamonds, and long-dashed with squares
denote the cases of $\gamma=3, 2.5, 2.25$ and 2, respectively.
Thick lines with filled symbols are for star 1 (less massive star)
and thin lines with open symbols for star 2 (massive star).
}
\end{figure}

In Fig.~\ref{fig:turning}, the quantity $\chi$
at the turning point of total angular momentum is shown
as a function of the mass ratio $M_1/M_2$ in the synchronized case
with $\gamma=3, 2.5, 2.25$ and 2.
It appears  clearly that the $\gamma=3$ curve has a minimum
at $M_1/M_2 \sim 0.2-0.3$.
This means that there always exists a turning point of
total angular momentum for synchronized binaries with $\gamma=3$.
One can also see that the turning point exists marginally
in the $\gamma=2.5$ case at around $M_1/M_2 \sim 0.2-0.3$.
On the other hand, it seems that the line of $\gamma=2.25$ may reach
$\chi=0$ (mass shedding limit) around $0.1<M_1/M_2<0.5$.
if we extrapolate it.
This means that there does not exist any turning point
for synchronized binaries with $\gamma \le 2.25$ and $0.1<M_1/M_2<0.5$.

Therefore we can draw two conclusions from Fig.~\ref{fig:turning}.
The first one is that it becomes difficult to find the turning point
for small $\gamma$.
The second one is that the mass ratio for which it is more
difficult to obtain any turning point is about $M_1/M_2 \sim 0.2-0.3$.
This can be explained as follows.
For equal mass binaries, the orbital motions of the two stars contribute
to the orbital angular momentum, which decreases
when the separation decreases.
In addition, the spins of the two stars contribute
to the spin angular momentum, which increases when the separation decreases.
The increase of the spin angular momentum has two parts.
The first one is the increase of orbital angular velocity,
and the second one comes from the increase of the moment of inertia.
Then, the total angular momentum has a minimum somewhere.
Keeping this fact in mind, let us consider the different mass case,
for example $M_1/M_2=0.5$.
There still exists large contribution from the massive star
to the orbital angular momentum and spin angular momentum,
while the moment of inertia of massive star does not increase so much
because its deformation is small.
Then, since the less massive star has to produce much more moment of inertia
in order to have the turning point,
the deformation at the turning point becomes large
and $\chi$ becomes small
(see Table~\ref{table1}, as well as Table I of Paper II).
This tendency continues up to $M_1/M_2 \sim 0.3$.
However, when the mass ratio becomes smaller than 0.2,
the situation changes.
Then the center of mass of the binary system
is located inside the massive star near the turning point.
Hence the contribution from the massive star
to the orbital angular momentum becomes small.
In addition, there is a large contribution from the massive star
to the spin angular momentum
because of the synchronous rotation.
Accordingly, the less massive star does not need to produce
large moment of inertia.
Therefore, it becomes easier again for the turning point to appear.

If we apply this explanation to the case of irrotational binaries,
we can understand why there does not appear any turning point
in the irrotational cases of different mass binaries.
First of all, let us recall the case of equal mass binaries.
The orbital motions of the two stars contribute
to the orbital angular momentum which decreases
when the separation decreases.
Additionally, the spins of the two stars contribute
to the spin angular momentum which increases when the separation decreases.
These situations are the same as in the synchronized case.
However, the spin angular momentum is much smaller than
in the synchronized case because of irrotation.
Next, we consider the different mass case.
The large difference from the synchronized case is the contribution
from the massive star to the spin angular momentum.
In the irrotational case, it becomes negligible when the mass difference
is large, in which case the massive star does not
deviate from a spherical body so much.
Then, in order to obtain the turning point,
the less massive star must produce much more moment of inertia,
i.e., much more deformation.
However, before obtaining the turning point,
the less massive star reaches the mass-shedding point.

\begin{figure}
\vspace{0.3cm}
  \centerline{ \epsfig{figure=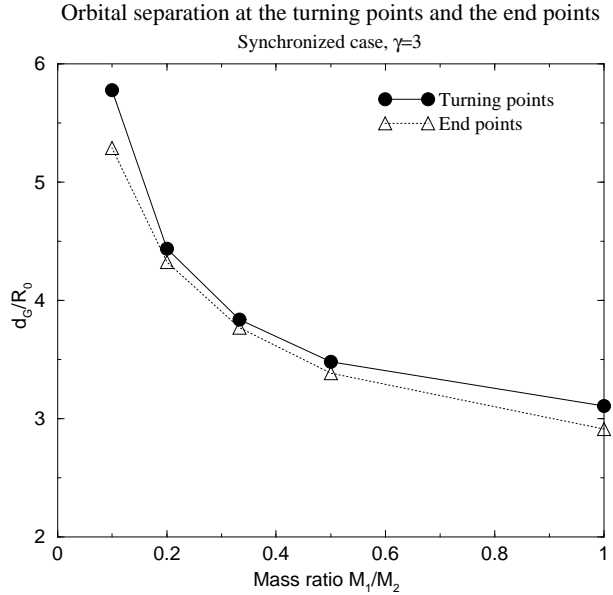,height=8cm} }
\vspace{0.3cm}
\caption[]{\label{fig:turning_end}
Orbital separation at the turning point of total angular momentum
and at the end point of a constant-mass sequence
as a function of the mass ratio $M_1/M_2$ in the synchronized case
with $\gamma=3$.
Solid line with filled circles and dotted one with open triangles
denote the turning point and the end point (cusp point), respectively.
}
\end{figure}

In Fig. \ref{fig:turning_end}, we show the orbital separation
at the turning point of total angular momentum
and at the end point of a constant-mass sequence
as a function of the  mass ratio $M_1/M_2$ in the synchronized case
with $\gamma=3$.
One can see from this figure that the orbital separation
(normalized by a radius of a spherical static star with the same mass)
increases when the mass ratio decreases.
It also appears in Fig.\ref{fig:turning_end} that when the mass ratio decreases,
the separation between the turning point and the end point decreases
first and increases again. This is in accordance with the behavior discussed
above (difficulty of finding the turning point for intermediate mass ratios,
$M_1/M_2 \sim 0.2-0.3$).

\section{Summary}

We have studied Newtonian equilibrium sequences of both synchronized and irrotational
binary systems on circular orbits and
composed of different mass stars.

It is found that the equilibrium sequences for different mass binaries
always terminate by the mass-shedding limit
of the less massive star (cusp point),
regardless of the rotation state (synchronized or irrotational).
This mass-shedding limit occurs for a detached configuration. This
contrasts with the case of equal mass synchronized binaries,
for which sequences terminate by the contact between the two stars (Paper~II).
This is due to the fact that the tidal force from the massive companion star
is larger in the different mass case,
so that it is easier for the less massive star to be disrupted before contact.

Regarding the turning point of the total angular momentum (or total energy)
along an evolutionary sequence,
we have found that it is difficult to appear for small mass ratio $M_1/M_2$.
In fact, for irrotational binaries,
we did not find any turning point for $M_1/M_2 \le 0.5$
in the region $\chi>0.2$.
Of course, there remains the possibility to find a turning point
in the region of $0<\chi<0.2$,
but it is difficult for our method to calculate in that region
because of the Gibbs phenomenon.
For synchronized binaries,
on the other hand, we have found that the turning point is
difficult to get for mass ratios $M_1/M_2 \sim 0.2-0.3$,
but easier again for smaller mass ratios, $M_1/M_2 < 0.2$.

\acknowledgments

We would like to thank Koji Uryu for useful discussions.
The code development and the numerical computations have been performed
on SGI workstations purchased thanks to a special grant from the C.N.R.S.


\newpage
\begin{table}
\caption{
}
 \begin{center}
  \begin{tabular}{rccccccccccc}
  \multicolumn{12}{c}{Synchronized case~~~~~$M_1/M_2=0.5$} \\
  $d_G/R_0$&$d/(R_0+R_0')$&$d/(a_1+a_1')$&
  $\bar{\Omega}$&$\bar{J}$&$\bar{E}$&Virial error&
  $a_2/a_1$&$a_3/a_1$&$a_{1,{\rm opp}}/a_1$&
  $(\rho_c-\rho_{c0})/\rho_{c0}$&$\chi$ \\
   & & & & & & &$a_2'/a_1'$&$a_3'/a_1'$&$a_{1,{\rm opp}}'/a_1'$&
  $(\rho_c'-\rho_{c0}')/\rho_{c0}'$&$\chi'$  \\ \hline
  \multicolumn{12}{c}{$\gamma=3$} \\
    7.515&3.497&3.464&0.09711&3.267&-2.619&5.140(-6)&
                 0.9853&0.9787&0.9981&-1.989(-3)&0.9639 \\
    & & & & & & &0.9944&0.9894&0.9992&-1.467(-3)&0.9820 \\

    6.413&2.984&2.938&0.1232&3.054&-2.639&5.016(-6)&
                 0.9756&0.9653&0.9963&-3.286(-3)&0.9406 \\
    & & & & & & &0.9907&0.9827&0.9984&-2.384(-3)&0.9705 \\

    5.411&2.518&2.450&0.1591&2.856&-2.663&4.841(-6)&
                 0.9574&0.9408&0.9921&-5.741(-3)&0.8975 \\
    & & & & & & &0.9839&0.9706&0.9968&-4.040(-3)&0.9494 \\

    4.307&2.005&1.884&0.2247&2.648&-2.697&4.479(-6)&
                 0.9047&0.8743&0.9758&-1.298(-2)&0.7762 \\
    & & & & & & &0.9649&0.9391&0.9910&-8.405(-3)&0.8926 \\

    4.006&1.865&1.716&0.2512&2.599&-2.707&4.322(-6)&
                 0.8729&0.8369&0.9633&-1.740(-2)&0.7044 \\
    & & & & & & &0.9541&0.9224&0.9871&-1.074(-2)&0.8614 \\

    3.702&1.725&1.530&0.2839&2.561&-2.716&4.121(-6)&
                 0.8190&0.7769&0.9371&-2.486(-2)&0.5817 \\
    & & & & & & &0.9373&0.8977&0.9802&-1.418(-2)&0.8137 \\

$\dagger$3.481&1.625&1.364&0.3135&2.548&-2.719&4.063(-6)&
                 0.7416&0.6962&0.8872&-3.493(-2)&0.3937 \\
    & & & & & & &0.9173&0.8704&0.9710&-1.794(-2)&0.7581 \\

    3.385&1.584&1.254&0.3288&2.554&-2.719&2.294(-4)&
                 0.6662&0.6221&0.8208&-4.251(-2)&0.2007 \\
    & & & & & & &0.9041&0.8534&0.9641&-2.018(-2)&0.7213 \\ \hline
  \multicolumn{12}{c}{$\gamma=2$} \\
    7.000&3.500&3.467&0.1080&3.129&-2.639&4.826(-13)&
                 0.9852&0.9787&0.9978&-4.216(-3)&0.9497 \\
    & & & & & & &0.9964&0.9930&0.9995&-2.048(-3)&0.9836 \\

    5.979&2.990&2.943&0.1368&2.918&-2.662&2.487(-12)&
                 0.9754&0.9652&0.9957&-6.930(-3)&0.9174 \\
    & & & & & & &0.9940&0.9887&0.9990&-3.307(-3)&0.9733 \\

    5.007&2.504&2.433&0.1787&2.710&-2.690&3.529(-13)&
                 0.9557&0.9388&0.9904&-1.237(-2)&0.8534 \\
    & & & & & & &0.9894&0.9804&0.9979&-5.698(-3)&0.9534 \\

    3.985&1.993&1.864&0.2522&2.491&-2.730&7.998(-11)&
                 0.8976&0.8668&0.9687&-2.778(-2)&0.6750 \\
    & & & & & & &0.9772&0.9596&0.9941&-1.165(-2)&0.9025 \\

    3.644&1.823&1.649&0.2891&2.424&-2.746&4.916(-10)&
                 0.8485&0.8107&0.9436&-3.988(-2)&0.5293 \\
    & & & & & & &0.9685&0.9456&0.9910&-1.559(-2)&0.8676 \\

    3.399&1.701&1.459&0.3220&2.383&-2.756&9.025(-8)&
                 0.7739&0.7318&0.8921&-5.505(-2)&0.3084 \\
    & & & & & & &0.9588&0.9307&0.9871&-1.970(-2)&0.8298 \\

    3.349&1.677&1.407&0.3295&2.376&-2.758&1.181(-5)&
                 0.7447&0.7024&0.8669&-5.948(-2)&0.2340 \\
    & & & & & & &0.9563&0.9270&0.9860&-2.073(-2)&0.8201 \\

    3.314&1.660&1.355&0.3350&2.372&-2.759&2.142(-5)&
                 0.7098&0.6682&0.8327&-6.308(-2)&0.1277 \\
    & & & & & & &0.9543&0.9241&0.9851&-2.150(-2)&0.8126 \\ \hline
  \multicolumn{12}{c}{$\gamma=1.8$} \\
    6.701&3.516&3.481&0.1153&3.051&-2.674&1.860(-9)&
                 0.9844&0.9777&0.9975&-5.678(-3)&0.9403 \\
    & & & & & & &0.9972&0.9946&0.9996&-2.046(-3)&0.9857 \\

    5.723&3.003&2.953&0.1461&2.842&-2.697&1.821(-9)&
                 0.9740&0.9634&0.9950&-9.329(-3)&0.9017 \\
    & & & & & & &0.9954&0.9913&0.9992&-3.298(-3)&0.9767 \\

    4.746&2.491&2.412&0.1936&2.622&-2.729&1.744(-9)&
                 0.9510&0.9329&0.9882&-1.719(-2)&0.8192 \\
    & & & & & & &0.9916&0.9844&0.9983&-5.839(-3)&0.9582 \\

    3.824&2.007&1.863&0.2681&2.413&-2.771&1.654(-9)&
                 0.8884&0.8566&0.9618&-3.703(-2)&0.6085 \\
    & & & & & & &0.9828&0.9691&0.9957&-1.142(-2)&0.9163 \\

    3.488&1.831&1.626&0.3085&2.341&-2.788&2.534(-8)&
                 0.8251&0.7865&0.9236&-5.377(-2)&0.4079 \\
    & & & & & & &0.9762&0.9582&0.9934&-1.531(-2)&0.8861 \\

    3.347&1.758&1.493&0.3286&2.313&-2.796&1.509(-6)&
                 0.7666&0.7261&0.8766&-6.483(-2)&0.2401 \\
    & & & & & & &0.9723&0.9520&0.9919&-1.751(-2)&0.8686 \\

    3.319&1.743&1.456&0.3330&2.308&-2.797&4.688(-6)&
                 0.7449&0.7045&0.8563&-6.754(-2)&0.1788 \\
    & & & & & & &0.9714&0.9505&0.9915&-1.800(-2)&0.8646 \\

    3.305&1.736&1.431&0.3352&2.306&-2.798&4.486(-6)&
                 0.7290&0.6889&0.8402&-6.906(-2)&0.1300 \\
    & & & & & & &0.9709&0.9498&0.9913&-1.826(-2)&0.8625 \\
  \end{tabular}
 \end{center}
 \label{table1}
\end{table}%

\begin{table}
\caption{
}
 \begin{center}
  \begin{tabular}{rccccccccccc}
  \multicolumn{12}{c}{Synchronized case~~~~~$M_1/M_2=0.2$} \\
  $d_G/R_0$&$d/(R_0+R_0')$&$d/(a_1+a_1')$&
  $\bar{\Omega}$&$\bar{J}$&$\bar{E}$&Virial error&
  $a_2/a_1$&$a_3/a_1$&$a_{1,{\rm opp}}/a_1$&
  $(\rho_c-\rho_{c0})/\rho_{c0}$&$\chi$ \\
   & & & & & & &$a_2'/a_1'$&$a_3'/a_1'$&$a_{1,{\rm opp}}'/a_1'$&
  $(\rho_c'-\rho_{c0}')/\rho_{c0}'$&$\chi'$  \\ \hline
  \multicolumn{12}{c}{$\gamma=3$} \\
    9.508&3.995&3.963&0.09649&6.583&-10.873&5.413(-6)&
                 0.9821&0.9757&0.9982&-1.992(-3)&0.9591 \\
    & & & & & & &0.9981&0.9946&0.9997&-9.960(-4)&0.9910 \\

    8.374&3.519&3.476&0.1168&6.257&-10.903&5.358(-6)&
                 0.9733&0.9641&0.9968&-2.993(-3)&0.9392 \\
    & & & & & & &0.9972&0.9921&0.9996&-1.463(-3)&0.9867 \\

    7.153&3.006&2.945&0.1480&5.906&-10.943&5.271(-6)&
                 0.9556&0.9413&0.9937&-5.057(-3)&0.9000 \\
    & & & & & & &0.9953&0.9872&0.9992&-2.362(-3)&0.9784 \\

    5.931&2.493&2.397&0.1962&5.573&-10.992&5.118(-6)&
                 0.9162&0.8928&0.9846&-9.892(-3)&0.8152 \\
    & & & & & & &0.9914&0.9772&0.9982&-4.201(-3)&0.9611 \\

    5.232&2.199&2.065&0.2372&5.409&-11.022&4.976(-6)&
                 0.8669&0.8355&0.9694&-1.639(-2)&0.7116 \\
    & & & & & & &0.9869&0.9661&0.9968&-6.220(-3)&0.9418 \\

    4.705&1.979&1.784&0.2790&5.321&-11.042&4.815(-6)&
                 0.7900&0.7520&0.9370&-2.715(-2)&0.5512 \\
    & & & & & & &0.9808&0.9523&0.9947&-8.749(-3)&0.9171 \\

$\dagger$4.437&1.869&1.605&0.3059&5.303&-11.046&4.486(-6)&
                 0.7075&0.6678&0.8869&-3.873(-2)&0.3683 \\
    & & & & & & &0.9759&0.9418&0.9928&-1.064(-2)&0.8979 \\

    4.324&1.825&1.481&0.3190&5.309&-11.042&1.142(-4)&
                 0.6273&0.5897&0.8166&-4.791(-2)&0.1834 \\
    & & & & & & &0.9731&0.9362&0.9917&-1.164(-2)&0.8874 \\ \hline
  \multicolumn{12}{c}{$\gamma=2$} \\
    8.021&4.011&3.962&0.1245&5.952&-13.302&3.428(-13)&
                 0.9754&0.9670&0.9967&-5.778(-3)&0.9225 \\
    & & & & & & &0.9990&0.9973&0.9999&-1.083(-3)&0.9937 \\

    7.000&3.500&3.433&0.1528&5.611&-13.343&1.012(-12)&
                 0.9618&0.9494&0.9940&-9.010(-3)&0.8809 \\
    & & & & & & &0.9985&0.9959&0.9998&-1.632(-3)&0.9904 \\

    5.979&2.990&2.892&0.1936&5.261&-13.395&2.583(-13)&
                 0.9350&0.9160&0.9875&-1.547(-2)&0.8017 \\
    & & & & & & &0.9976&0.9933&0.9996&-2.629(-3)&0.9844 \\

    5.347&2.674&2.541&0.2291&5.043&-13.435&5.889(-12)&
                 0.9029&0.8775&0.9776&-2.334(-2)&0.7098 \\
    & & & & & & &0.9966&0.9906&0.9994&-3.694(-3)&0.9780 \\

    5.006&2.504&2.342&0.2530&4.927&-13.459&9.100(-11)&
                 0.8745&0.8450&0.9670&-3.029(-2)&0.6310 \\
    & & & & & & &0.9958&0.9885&0.9992&-4.517(-3)&0.9729 \\

    4.616&2.309&2.091&0.2862&4.800&-13.489&6.401(-9)&
                 0.8210&0.7862&0.9418&-4.309(-2)&0.4852 \\
    & & & & & & &0.9945&0.9852&0.9988&-5.800(-3)&0.9650 \\

    4.371&2.188&1.898&0.3111&4.727&-13.508&2.342(-7)&
                 0.7577&0.7203&0.9011&-5.652(-2)&0.3128 \\
    & & & & & & &0.9933&0.9824&0.9985&-6.876(-3)&0.9583 \\

    4.247&2.127&1.747&0.3252&4.694&-13.517&2.828(-6)&
                 0.6868&0.6500&0.8381&-6.673(-2)&0.1319 \\
    & & & & & & &0.9926&0.9806&0.9983&-7.529(-3)&0.9541 \\ \hline
  \multicolumn{12}{c}{$\gamma=1.8$} \\
    7.182&4.002&3.931&0.1470&5.595&-15.489&1.937(-9)&
                 0.9678&0.9572&0.9950&-9.643(-3)&0.8864 \\
    & & & & & & &0.9994&0.9983&0.9999&-8.933(-4)&0.9954 \\

    6.288&3.504&3.406&0.1794&5.272&-15.535&1.919(-9)&
                 0.9499&0.9345&0.9907&-1.499(-2)&0.8263 \\
    & & & & & & &0.9991&0.9974&0.9999&-1.333(-3)&0.9932 \\

    5.394&3.006&2.859&0.2260&4.937&-15.593&1.897(-9)&
                 0.9138&0.8907&0.9797&-2.573(-2)&0.7103 \\
    & & & & & & &0.9985&0.9959&0.9998&-2.118(-3)&0.9890 \\

    4.931&2.748&2.555&0.2587&4.761&-15.630&1.731(-9)&
                 0.8776&0.8489&0.9656&-3.625(-2)&0.5992 \\
    & & & & & & &0.9980&0.9945&0.9997&-2.780(-3)&0.9856 \\

    4.499&2.508&2.229&0.2972&4.598&-15.669&2.131(-8)&
                 0.8128&0.7783&0.9313&-5.346(-2)&0.4088 \\
    & & & & & & &0.9974&0.9928&0.9995&-3.678(-3)&0.9808 \\

    4.313&2.405&2.046&0.3170&4.531&-15.687&3.586(-7)&
                 0.7555&0.7194&0.8896&-6.544(-2)&0.2571 \\
    & & & & & & &0.9970&0.9917&0.9994&-4.188(-3)&0.9780 \\

    4.281&2.388&2.006&0.3205&4.520&-15.690&2.158(-7)&
                 0.7400&0.7039&0.8762&-6.792(-2)&0.2194 \\
    & & & & & & &0.9969&0.9915&0.9994&-4.283(-3)&0.9775 \\

    4.241&2.365&1.938&0.3252&4.506&-15.694&7.897(-7)&
                 0.7113&0.6756&0.8487&-7.146(-2)&0.1414 \\
    & & & & & & &0.9968&0.9913&0.9994&-4.411(-3)&0.9768 \\
  \end{tabular}
 \end{center}
 \label{table2}
\end{table}%

\begin{table}
\caption{
}
 \begin{center}
  \begin{tabular}{rccccccccccc}
  \multicolumn{12}{c}{Synchronized case~~~~~$M_1/M_2=0.1$} \\
  $d_G/R_0$&$d/(R_0+R_0')$&$d/(a_1+a_1')$&
  $\bar{\Omega}$&$\bar{J}$&$\bar{E}$&Virial error&
  $a_2/a_1$&$a_3/a_1$&$a_{1,{\rm opp}}/a_1$&
  $(\rho_c-\rho_{c0})/\rho_{c0}$&$\chi$ \\
   & & & & & & &$a_2'/a_1'$&$a_3'/a_1'$&$a_{1,{\rm opp}}'/a_1'$&
  $(\rho_c'-\rho_{c0}')/\rho_{c0}'$&$\chi'$  \\ \hline
  \multicolumn{12}{c}{$\gamma=3$} \\
    12.926&5.000&4.974&0.08242&11.450&-35.974&5.562(-6)&
                 0.9861&0.9814&0.9989&-1.439(-3)&0.9689 \\
    & & & & & & &0.9994&0.9975&0.9999&-5.488(-4)&0.9959 \\

    10.320&3.993&3.950&0.1155&10.543&-36.051&5.510(-6)&
                 0.9718&0.9628&0.9973&-2.957(-3)&0.9376 \\
    & & & & & & &0.9989&0.9951&0.9998&-1.081(-3)&0.9918 \\

    9.018&3.489&3.431&0.1415&10.107&-36.099&5.466(-6)&
                 0.9566&0.9435&0.9950&-4.644(-3)&0.9049 \\
    & & & & & & &0.9983&0.9926&0.9997&-1.626(-3)&0.9877 \\

    7.815&3.024&2.942&0.1754&9.741&-36.148&5.402(-6)&
                 0.9304&0.9113&0.9904&-7.708(-3)&0.8497 \\
    & & & & & & &0.9973&0.9886&0.9995&-2.512(-3)&0.9809 \\

    6.511&2.520&2.388&0.2310&9.442&-36.200&5.276(-6)&
                 0.8669&0.8373&0.9750&-1.589(-2)&0.7207 \\
    & & & & & & &0.9950&0.9800&0.9989&-4.398(-3)&0.9663 \\

$\dagger$5.776&2.237&2.037&0.2771&9.376&-36.214&5.158(-6)&
                 0.7805&0.7440&0.9436&-2.839(-2)&0.5491 \\
    & & & & & & &0.9925&0.9708&0.9981&-6.395(-3)&0.9506 \\

    5.397&2.093&1.801&0.3078&9.407&-36.206&6.572(-6)&
                 0.6769&0.6394&0.8835&-4.404(-2)&0.3379 \\
    & & & & & & &0.9904&0.9637&0.9974&-7.957(-3)&0.9382 \\

    5.289&2.054&1.687&0.3179&9.434&-36.201&2.323(-5)&
                 0.6087&0.5733&0.8218&-5.190(-2)&0.1899 \\
    & & & & & & &0.9896&0.9611&0.9971&-8.510(-3)&0.9337 \\ \hline
  \multicolumn{12}{c}{$\gamma=2$} \\
    10.014&5.007&4.949&0.1209&9.844&-50.983&4.219(-13)&
                 0.9751&0.9672&0.9973&-5.466(-3)&0.9236 \\
    & & & & & & &0.9998&0.9989&1.0000&-5.090(-4)&0.9975 \\

    8.021&4.011&3.914&0.1686&8.964&-51.092&7.547(-13)&
                 0.9491&0.9343&0.9928&-1.143(-2)&0.8468 \\
    & & & & & & &0.9995&0.9979&0.9999&-9.922(-4)&0.9951 \\

    7.000&3.500&3.364&0.2069&8.501&-51.167&1.083(-13)&
                 0.9189&0.8977&0.9859&-1.859(-2)&0.7616 \\
    & & & & & & &0.9993&0.9968&0.9999&-1.496(-3)&0.9927 \\

    5.978&2.990&2.771&0.2624&8.046&-51.259&1.180(-10)&
                 0.8525&0.8220&0.9648&-3.492(-2)&0.5836 \\
    & & & & & & &0.9988&0.9949&0.9998&-2.412(-3)&0.9881 \\

    5.588&2.795&2.510&0.2907&7.882&-51.298&4.957(-9)&
                 0.8001&0.7657&0.9412&-4.772(-2)&0.4475 \\
    & & & & & & &0.9985&0.9937&0.9997&-2.962(-3)&0.9853 \\

    5.343&2.674&2.308&0.3112&7.786&-51.323&9.742(-8)&
                 0.7422&0.7061&0.9055&-6.047(-2)&0.2982 \\
    & & & & & & &0.9983&0.9927&0.9997&-3.399(-3)&0.9831 \\

    5.244&2.625&2.198&0.3202&7.749&-51.333&4.620(-7)&
                 0.7021&0.6662&0.8735&-6.753(-2)&0.2108 \\
    & & & & & & &0.9982&0.9923&0.9997&-3.601(-3)&0.9821 \\

    5.210&2.608&2.145&0.3235&7.737&-51.336&8.403(-7)&
                 0.6802&0.6448&0.8528&-7.057(-2)&0.1531 \\
    & & & & & & &0.9981&0.9922&0.9996&-3.676(-3)&0.9817 \\ \hline
  \multicolumn{12}{c}{$\gamma=1.8$} \\
    8.599&5.000&4.898&0.1519&9.032&-65.879&1.971(-9)&
                 0.9629&0.9517&0.9950&-1.048(-2)&0.8728 \\
    & & & & & & &0.9999&0.9994&1.0000&-3.540(-4)&0.9985 \\

    6.868&3.994&3.814&0.2129&8.173&-66.013&1.962(-9)&
                 0.9204&0.8998&0.9853&-2.278(-2)&0.7380 \\
    & & & & & & &0.9997&0.9989&1.0000&-6.959(-4)&0.9970 \\

    6.030&3.507&3.242&0.2589&7.739&-66.101&1.903(-9)&
                 0.8696&0.8414&0.9686&-3.759(-2)&0.5869 \\
    & & & & & & &0.9996&0.9983&1.0000&-1.030(-3)&0.9956 \\

    5.862&3.409&3.117&0.2702&7.651&-66.121&1.753(-9)&
                 0.8531&0.8232&0.9619&-4.228(-2)&0.5397 \\
    & & & & & & &0.9996&0.9982&0.9999&-1.122(-3)&0.9952 \\

    5.582&3.247&2.891&0.2909&7.506&-66.156&1.359(-9)&
                 0.8159&0.7831&0.9440&-5.245(-2)&0.4354 \\
    & & & & & & &0.9995&0.9979&0.9999&-1.301(-3)&0.9944 \\

    5.302&3.085&2.610&0.3145&7.362&-66.194&1.360(-7)&
                 0.7511&0.7162&0.9018&-6.743(-2)&0.2589 \\
    & & & & & & &0.9994&0.9975&0.9999&-1.521(-3)&0.9934 \\

    5.246&3.052&2.535&0.3197&7.334&-66.202&1.308(-7)&
                 0.7294&0.6945&0.8843&-7.140(-2)&0.2193 \\
    & & & & & & &0.9994&0.9974&0.9999&-1.572(-3)&0.9932 \\

    5.203&3.028&2.463&0.3236&7.313&-66.207&1.021(-7)&
                 0.7064&0.6718&0.8630&-7.466(-2)&0.1599 \\
    & & & & & & &0.9994&0.9974&0.9999&-1.611(-3)&0.9930 \\
  \end{tabular}
 \end{center}
 \label{table3}
\end{table}%

\begin{table}
\caption{
}
 \begin{center}
  \begin{tabular}{rccccccccccc}
  \multicolumn{12}{c}{Irrotational case~~~~~$M_1/M_2=0.5$} \\
  $d_G/R_0$&$d/(R_0+R_0')$&$d/(a_1+a_1')$&
  $\bar{\Omega}$&$\bar{J}$&$\bar{E}$&Virial error&
  $a_2/a_1$&$a_3/a_1$&$a_{1,{\rm opp}}/a_1$&
  $(\rho_c-\rho_{c0})/\rho_{c0}$&$\chi$ \\
   & & & & & & &$a_2'/a_1'$&$a_3'/a_1'$&$a_{1,{\rm opp}}'/a_1'$&
  $(\rho_c'-\rho_{c0}')/\rho_{c0}'$&$\chi'$  \\ \hline
  \multicolumn{12}{c}{$\gamma=3$} \\
    7.515&3.497&3.473&0.09710&3.166&-2.623&5.208(-6)&
                 0.9852&0.9854&0.9981&-4.380(-5)&0.9746 \\
    & & & & & & &0.9944&0.9944&0.9992&-6.207(-6)&0.9902 \\

    6.413&2.984&2.950&0.1232&2.925&-2.646&5.124(-6)&
                 0.9754&0.9758&0.9963&-1.174(-4)&0.9576 \\
    & & & & & & &0.9907&0.9908&0.9984&-1.647(-5)&0.9837 \\

    5.411&2.518&2.467&0.1591&2.689&-2.675&5.014(-6)&
                 0.9567&0.9580&0.9921&-3.470(-4)&0.9250 \\
    & & & & & & &0.9837&0.9841&0.9968&-4.769(-5)&0.9712 \\

    4.308&2.005&1.910&0.2245&2.406&-2.721&4.811(-6)&
                 0.9015&0.9066&0.9755&-1.643(-3)&0.8257 \\
    & & & & & & &0.9641&0.9656&0.9909&-2.116(-4)&0.9349 \\

    4.006&1.865&1.745&0.2508&2.325&-2.738&4.724(-6)&
                 0.8676&0.8756&0.9624&-2.846(-3)&0.7611 \\
    & & & & & & &0.9528&0.9552&0.9869&-3.507(-4)&0.9132 \\

    3.703&1.725&1.562&0.2832&2.245&-2.756&4.595(-6)&
                 0.8089&0.8219&0.9343&-5.529(-3)&0.6394 \\
    & & & & & & &0.9350&0.9389&0.9799&-6.280(-4)&0.8778 \\

    3.498&1.632&1.408&0.3101&2.197&-2.768&2.786(-6)&
                 0.7318&0.7501&0.8855&-9.876(-3)&0.4528 \\
    & & & & & & &0.9156&0.9214&0.9712&-1.001(-3)&0.8366 \\

    3.401&1.590&1.298&0.3250&2.180&-2.773&2.041(-5)&
                 0.6553&0.6762&0.8210&-1.437(-2)&0.2303 \\
    & & & & & & &0.9018&0.9089&0.9643&-1.300(-3)&0.8050 \\ \hline
  \multicolumn{12}{c}{$\gamma=2$} \\
    7.000&3.500&3.478&0.1080&3.055&-2.643&1.061(-12)&
                 0.9852&0.9853&0.9978&-8.155(-5)&0.9647 \\
    & & & & & & &0.9964&0.9964&0.9995&-4.977(-6)&0.9912 \\

    5.979&2.990&2.957&0.1368&2.824&-2.667&1.429(-12)&
                 0.9754&0.9758&0.9957&-2.165(-4)&0.9411 \\
    & & & & & & &0.9940&0.9941&0.9990&-1.301(-5)&0.9855 \\

    5.007&2.504&2.454&0.1786&2.586&-2.700&5.191(-13)&
                 0.9556&0.9567&0.9905&-6.673(-4)&0.8930 \\
    & & & & & & &0.9894&0.9895&0.9979&-3.885(-5)&0.9741 \\

    3.986&1.993&1.899&0.2520&2.311&-2.750&3.427(-10)&
                 0.8978&0.9023&0.9695&-3.108(-3)&0.7476 \\
    & & & & & & &0.9772&0.9778&0.9942&-1.647(-4)&0.9432 \\

    3.644&1.823&1.692&0.2887&2.214&-2.773&4.678(-8)&
                 0.8493&0.8570&0.9453&-6.072(-3)&0.6143 \\
    & & & & & & &0.9686&0.9696&0.9912&-2.975(-4)&0.9209 \\

    3.400&1.701&1.512&0.3212&2.145&-2.791&5.162(-6)&
                 0.7771&0.7888&0.8962&-1.091(-2)&0.3791 \\
    & & & & & & &0.9592&0.9607&0.9874&-4.805(-4)&0.8957 \\

    3.350&1.677&1.463&0.3286&2.132&-2.795&1.852(-5)&
                 0.7494&0.7622&0.8729&-1.255(-2)&0.2773 \\
    & & & & & & &0.9567&0.9584&0.9864&-5.338(-4)&0.8890 \\

    3.326&1.665&1.432&0.3324&2.125&-2.797&4.053(-5)&
                 0.7296&0.7431&0.8547&-1.355(-2)&0.2014 \\
    & & & & & & &0.9553&0.9571&0.9858&-5.638(-4)&0.8853 \\ \hline
  \multicolumn{12}{c}{$\gamma=1.8$} \\
    6.701&3.516&3.493&0.1153&2.989&-2.677&1.871(-9)&
                 0.9844&0.9846&0.9975&-1.045(-4)&0.9580 \\
    & & & & & & &0.9972&0.9972&0.9996&-3.504(-6)&0.9924 \\

    5.723&3.003&2.970&0.1461&2.763&-2.702&1.857(-9)&
                 0.9741&0.9745&0.9951&-2.774(-4)&0.9299 \\
    & & & & & & &0.9954&0.9954&0.9992&-9.128(-6)&0.9874 \\

    4.746&2.491&2.437&0.1935&2.517&-2.738&1.811(-9)&
                 0.9515&0.9527&0.9885&-9.087(-4)&0.8678 \\
    & & & & & & &0.9916&0.9917&0.9983&-2.874(-5)&0.9769 \\

    3.824&2.007&1.905&0.2680&2.263&-2.788&5.834(-10)&
                 0.8913&0.8957&0.9638&-3.888(-3)&0.6959 \\
    & & & & & & &0.9829&0.9832&0.9958&-1.109(-4)&0.9523 \\

    3.489&1.831&1.682&0.3080&2.165&-2.813&5.413(-7)&
                 0.8327&0.8402&0.9291&-7.716(-3)&0.5055 \\
    & & & & & & &0.9765&0.9771&0.9936&-2.006(-4)&0.9337 \\

    3.348&1.758&1.559&0.3280&2.124&-2.824&1.519(-5)&
                 0.7783&0.7883&0.8856&-1.087(-2)&0.2930 \\
    & & & & & & &0.9727&0.9735&0.9922&-2.636(-4)&0.9227 \\

    3.334&1.751&1.542&0.3301&2.120&-2.825&2.569(-5)&
                 0.7688&0.7791&0.8768&-1.131(-2)&0.2534 \\
    & & & & & & &0.9723&0.9731&0.9920&-2.714(-4)&0.9214 \\

    3.323&1.745&1.527&0.3318&2.116&-2.826&3.720(-5)&
                 0.7596&0.7702&0.8679&-1.170(-2)&0.2149 \\
    & & & & & & &0.9720&0.9727&0.9919&-2.777(-4)&0.9203 \\
  \end{tabular}
 \end{center}
 \label{table4}
\end{table}%

\begin{table}
\caption{
}
 \begin{center}
  \begin{tabular}{rccccccccccc}
  \multicolumn{12}{c}{Irrotational case~~~~~$M_1/M_2=0.2$} \\
  $d_G/R_0$&$d/(R_0+R_0')$&$d/(a_1+a_1')$&
  $\bar{\Omega}$&$\bar{J}$&$\bar{E}$&Virial error&
  $a_2/a_1$&$a_3/a_1$&$a_{1,{\rm opp}}/a_1$&
  $(\rho_c-\rho_{c0})/\rho_{c0}$&$\chi$ \\
   & & & & & & &$a_2'/a_1'$&$a_3'/a_1'$&$a_{1,{\rm opp}}'/a_1'$&
  $(\rho_c'-\rho_{c0}')/\rho_{c0}'$&$\chi'$  \\ \hline
  \multicolumn{12}{c}{$\gamma=3$} \\
    9.508&3.995&3.971&0.09649&6.295&-10.885&5.459(-6)&
                 0.9820&0.9822&0.9982&-6.669(-5)&0.9696 \\
    & & & & & & &0.9981&0.9981&0.9997&-7.198(-7)&0.9967 \\

    8.374&3.519&3.487&0.1167&5.908&-10.920&5.427(-6)&
                 0.9730&0.9735&0.9968&-1.467(-4)&0.9544 \\
    & & & & & & &0.9972&0.9972&0.9996&-1.560(-6)&0.9950 \\

    7.153&3.006&2.960&0.1479&5.462&-10.971&5.380(-6)&
                 0.9549&0.9561&0.9936&-3.981(-4)&0.9237 \\
    & & & & & & &0.9953&0.9954&0.9992&-4.107(-6)&0.9917 \\

    5.931&2.493&2.418&0.1961&4.978&-11.043&5.311(-6)&
                 0.9140&0.9177&0.9844&-1.380(-3)&0.8539 \\
    & & & & & & &0.9913&0.9915&0.9981&-1.327(-5)&0.9844 \\

    5.232&2.199&2.090&0.2370&4.683&-11.098&5.255(-6)&
                 0.8616&0.8694&0.9687&-3.415(-3)&0.7621 \\
    & & & & & & &0.9866&0.9871&0.9968&-2.981(-5)&0.9757 \\

    4.705&1.979&1.813&0.2786&4.456&-11.148&5.179(-6)&
                 0.7777&0.7926&0.9337&-8.255(-3)&0.6022 \\
    & & & & & & &0.9803&0.9812&0.9947&-6.105(-5)&0.9636 \\

    4.437&1.869&1.627&0.3054&4.347&-11.176&3.723(-6)&
                 0.6847&0.7054&0.8765&-1.529(-2)&0.3933 \\
    & & & & & & &0.9752&0.9764&0.9928&-9.337(-5)&0.9535 \\

    4.362&1.840&1.551&0.3138&4.321&-11.183&3.632(-6)&
                 0.6343&0.6565&0.8346&-1.928(-2)&0.2596 \\
    & & & & & & &0.9733&0.9747&0.9920&-1.066(-4)&0.9496 \\ \hline
  \multicolumn{12}{c}{$\gamma=2$} \\
    8.021&4.011&3.975&0.1245&5.782&-13.312&7.577(-13)&
                 0.9754&0.9757&0.9968&-2.289(-4)&0.9421 \\
    & & & & & & &0.9990&0.9990&0.9999&-3.480(-7)&0.9977 \\

    7.000&3.500&3.451&0.1528&5.402&-13.357&1.708(-14)&
                 0.9617&0.9625&0.9941&-5.369(-4)&0.9100 \\
    & & & & & & &0.9985&0.9985&0.9998&-7.936(-7)&0.9965 \\

    5.979&2.990&2.916&0.1936&4.994&-13.418&1.624(-12)&
                 0.9349&0.9369&0.9877&-1.482(-3)&0.8468 \\
    & & & & & & &0.9976&0.9976&0.9996&-2.072(-6)&0.9942 \\

    5.347&2.674&2.572&0.2290&4.726&-13.467&1.954(-11)&
                 0.9028&0.9066&0.9780&-3.147(-3)&0.7703 \\
    & & & & & & &0.9966&0.9966&0.9994&-4.116(-6)&0.9917 \\

    5.006&2.504&2.378&0.2529&4.575&-13.498&4.254(-10)&
                 0.8746&0.8802&0.9677&-5.018(-3)&0.7016 \\
    & & & & & & &0.9958&0.9958&0.9992&-6.190(-6)&0.9896 \\

    4.616&2.309&2.136&0.2859&4.399&-13.539&3.578(-8)&
                 0.8217&0.8308&0.9433&-9.292(-3)&0.5645 \\
    & & & & & & &0.9945&0.9946&0.9988&-1.029(-5)&0.9864 \\

    4.372&2.188&1.952&0.3107&4.287&-13.568&1.301(-6)&
                 0.7601&0.7726&0.9046&-1.474(-2)&0.3826 \\
    & & & & & & &0.9934&0.9935&0.9985&-1.458(-5)&0.9835 \\

    4.263&2.134&1.838&0.3229&4.238&-13.581&1.304(-5)&
                 0.7077&0.7220&0.8610&-1.878(-2)&0.2068 \\
    & & & & & & &0.9928&0.9929&0.9983&-1.719(-5)&0.9820 \\ \hline
  \multicolumn{12}{c}{$\gamma=1.8$} \\
    7.182&4.002&3.951&0.1470&5.471&-15.497&1.945(-9)&
                 0.9680&0.9685&0.9950&-4.441(-4)&0.9148 \\
    & & & & & & &0.9994&0.9994&0.9999&-1.670(-7)&0.9984 \\

    6.288&3.504&3.433&0.1794&5.120&-15.547&1.935(-9)&
                 0.9503&0.9514&0.9909&-1.029(-3)&0.8679 \\
    & & & & & & &0.9991&0.9991&0.9999&-3.727(-7)&0.9975 \\

    5.394&3.006&2.897&0.2259&4.745&-15.612&1.917(-9)&
                 0.9150&0.9178&0.9805&-2.805(-3)&0.7740 \\
    & & & & & & &0.9985&0.9985&0.9998&-9.451(-7)&0.9960 \\

    4.932&2.748&2.603&0.2586&4.539&-15.655&1.331(-9)&
                 0.8804&0.8851&0.9673&-5.203(-3)&0.6796 \\
    & & & & & & &0.9980&0.9981&0.9997&-1.634(-6)&0.9947 \\

    4.499&2.508&2.296&0.2970&4.340&-15.703&1.190(-7)&
                 0.8200&0.8280&0.9363&-1.026(-2)&0.5018 \\
    & & & & & & &0.9974&0.9974&0.9995&-2.877(-6)&0.9928 \\

    4.313&2.405&2.127&0.3167&4.253&-15.726&2.921(-6)&
                 0.7680&0.7783&0.8990&-1.447(-2)&0.3232 \\
    & & & & & & &0.9970&0.9970&0.9994&-3.729(-6)&0.9917 \\

    4.282&2.388&2.090&0.3202&4.239&-15.730&5.753(-6)&
                 0.7539&0.7647&0.8869&-1.542(-2)&0.2708 \\
    & & & & & & &0.9969&0.9969&0.9994&-3.934(-6)&0.9915 \\

    4.251&2.370&2.046&0.3237&4.224&-15.734&1.294(-5)&
                 0.7352&0.7466&0.8694&-1.651(-2)&0.1998 \\
    & & & & & & &0.9969&0.9969&0.9994&-4.105(-6)&0.9913 \\
  \end{tabular}
 \end{center}
 \label{table5}
\end{table}%

\begin{table}
\caption{
}
 \begin{center}
  \begin{tabular}{rccccccccccc}
  \multicolumn{12}{c}{Irrotational case~~~~~$M_1/M_2=0.1$} \\
  $d_G/R_0$&$d/(R_0+R_0')$&$d/(a_1+a_1')$&
  $\bar{\Omega}$&$\bar{J}$&$\bar{E}$&Virial error&
  $a_2/a_1$&$a_3/a_1$&$a_{1,{\rm opp}}/a_1$&
  $(\rho_c-\rho_{c0})/\rho_{c0}$&$\chi$ \\
   & & & & & & &$a_2'/a_1'$&$a_3'/a_1'$&$a_{1,{\rm opp}}'/a_1'$&
  $(\rho_c'-\rho_{c0}')/\rho_{c0}'$&$\chi'$  \\ \hline
  \multicolumn{12}{c}{$\gamma=3$} \\
    12.926&5.000&4.981&0.08241&10.841&-35.996&5.587(-6)&
                 0.9860&0.9861&0.9989&-4.162(-5)&0.9766 \\
    & & & & & & &0.9994&0.9994&0.9999&-6.530(-8)&0.9990 \\

    10.320&3.993&3.960&0.1155&9.688&-36.093&5.561(-6)&
                 0.9716&0.9720&0.9973&-1.671(-4)&0.9525 \\
    & & & & & & &0.9989&0.9989&0.9998&-2.540(-7)&0.9980 \\

    9.018&3.489&3.444&0.1415&9.057&-36.163&5.542(-6)&
                 0.9560&0.9571&0.9950&-3.917(-4)&0.9266 \\
    & & & & & & &0.9983&0.9983&0.9997&-5.776(-7)&0.9970 \\

    7.815&3.024&2.960&0.1754&8.434&-36.248&5.518(-6)&
                 0.9289&0.9315&0.9903&-9.961(-4)&0.8816 \\
    & & & & & & &0.9973&0.9973&0.9995&-1.393(-6)&0.9951 \\

    6.512&2.520&2.413&0.2309&7.708&-36.375&5.482(-6)&
                 0.8618&0.8694&0.9744&-3.594(-3)&0.7690 \\
    & & & & & & &0.9950&0.9951&0.9989&-4.379(-6)&0.9909 \\

    5.706&2.210&2.026&0.2821&7.236&-36.479&5.431(-6)&
                 0.7521&0.7690&0.9334&-1.087(-2)&0.5695 \\
    & & & & & & &0.9920&0.9923&0.9980&-1.037(-5)&0.9854 \\

    5.397&2.093&1.818&0.3075&7.063&-36.523&4.325(-6)&
                 0.6482&0.6710&0.8696&-2.031(-2)&0.3440 \\
    & & & & & & &0.9902&0.9906&0.9974&-1.524(-5)&0.9818 \\

    5.343&2.074&1.762&0.3124&7.036&-36.531&1.675(-6)&
                 0.6132&0.6367&0.8400&-2.362(-2)&0.2549 \\
    & & & & & & &0.9898&0.9902&0.9972&-1.641(-5)&0.9810 \\ \hline
  \multicolumn{12}{c}{$\gamma=2$} \\
    10.014&5.007&4.963&0.1209&9.542&-50.999&3.959(-13)&
                 0.9751&0.9755&0.9973&-2.407(-4)&0.9421 \\
    & & & & & & &0.9998&0.9998&1.0000&-2.281(-8)&0.9994 \\

    8.021&4.011&3.937&0.1686&8.542&-51.123&4.836(-13)&
                 0.9489&0.9502&0.9929&-9.720(-4)&0.8816 \\
    & & & & & & &0.9995&0.9995&0.9999&-8.702(-8)&0.9989 \\

    7.000&3.500&3.395&0.2069&7.982&-51.214&3.847(-14)&
                 0.9187&0.9215&0.9862&-2.370(-3)&0.8121 \\
    & & & & & & &0.9993&0.9993&0.9999&-1.986(-7)&0.9982 \\

    5.978&2.990&2.816&0.2624&7.383&-51.335&4.664(-10)&
                 0.8526&0.8597&0.9657&-7.186(-3)&0.6571 \\
    & & & & & & &0.9988&0.9988&0.9998&-5.205(-7)&0.9971 \\

    5.588&2.795&2.564&0.2905&7.144&-51.392&2.349(-8)&
                 0.8008&0.8114&0.9429&-1.218(-2)&0.5264 \\
    & & & & & & &0.9985&0.9985&0.9997&-7.896(-7)&0.9964 \\

    5.344&2.674&2.372&0.3109&6.993&-51.431&5.703(-7)&
                 0.7448&0.7584&0.9091&-1.797(-2)&0.3673 \\
    & & & & & & &0.9983&0.9983&0.9997&-1.044(-6)&0.9958 \\

    5.245&2.625&2.270&0.3198&6.932&-51.448&3.080(-6)&
                 0.7073&0.7223&0.8800&-2.150(-2)&0.2491 \\
    & & & & & & &0.9982&0.9982&0.9997&-1.175(-6)&0.9955 \\

    5.221&2.613&2.238&0.3221&6.917&-51.452&5.224(-6)&
                 0.6940&0.7094&0.8682&-2.257(-2)&0.2053 \\
    & & & & & & &0.9981&0.9982&0.9996&-1.211(-6)&0.9954 \\ \hline
  \multicolumn{12}{c}{$\gamma=1.8$} \\
    8.599&5.000&4.924&0.1519&8.843&-65.891&1.975(-9)&
                 0.9631&0.9637&0.9951&-6.061(-4)&0.9029 \\
    & & & & & & &0.9999&0.9999&1.0000&-7.780(-9)&0.9997 \\

    6.868&3.994&3.859&0.2129&7.905&-66.037&1.969(-9)&
                 0.9214&0.9238&0.9857&-2.565(-3)&0.7949 \\
    & & & & & & &0.9997&0.9997&1.0000&-3.015(-8)&0.9993 \\

    6.030&3.507&3.304&0.2589&7.411&-66.138&1.799(-9)&
                 0.8725&0.8777&0.9702&-6.246(-3)&0.6670 \\
    & & & & & & &0.9996&0.9996&1.0000&-6.629(-8)&0.9990 \\

    5.583&3.247&2.972&0.2908&7.135&-66.203&1.143(-8)&
                 0.8224&0.8304&0.9480&-1.103(-2)&0.5280 \\
    & & & & & & &0.9995&0.9995&0.9999&-1.060(-7)&0.9987 \\

    5.303&3.085&2.715&0.3143&6.959&-66.250&6.188(-7)&
                 0.7639&0.7747&0.9110&-1.671(-2)&0.3423 \\
    & & & & & & &0.9994&0.9994&0.9999&-1.460(-7)&0.9984 \\

    5.246&3.052&2.646&0.3194&6.923&-66.259&1.827(-6)&
                 0.7440&0.7556&0.8949&-1.834(-2)&0.2723 \\
    & & & & & & &0.9994&0.9994&0.9999&-1.533(-7)&0.9984 \\

    5.232&3.044&2.627&0.3207&6.914&-66.262&2.441(-6)&
                 0.7379&0.7497&0.8896&-1.879(-2)&0.2501 \\
    & & & & & & &0.9994&0.9994&0.9999&-1.583(-7)&0.9984 \\

    5.224&3.039&2.614&0.3215&6.909&-66.263&2.940(-6)&
                 0.7339&0.7458&0.8860&-1.908(-2)&0.2356 \\
    & & & & & & &0.9994&0.9994&0.9999&-1.598(-7)&0.9984 \\
  \end{tabular}
 \end{center}
 \label{table6}
\end{table}%

\end{document}